\documentclass[twocolumn,secnumarabic,amssymb, nobibnotes, aps, prd, showpacs]{revtex4}
\usepackage{amsmath,amsthm,amssymb}
\usepackage[dvips]{graphicx,color}
\begin{document}

\title{
Relations between allometric scalings and fluctuations in complex systems: The case of Japanese firms}
\author{Hayafumi Watanabe$^1$}\email[E-mail: ]{h-watanabe@smp.dis.titech.ac.jp}
\author{Hideki Takayasu$^2$}
\author{Misako Takayasu$^1$}

\affiliation{$^1$Department of Computational Intelligence and Systems Science, Interdisciplinary Graduate School of Science and Engineering, Tokyo Institute of Technology, 4259 Nagatsuta-cho, Midori-ku, Yokohama 226-8502, Japan}
\affiliation{$^2$Sony Computer Science Laboratories, 3-14-13 Higashi-Gotanda, Shinagawa-ku, Tokyo 141-0022, Japan}

\begin{abstract}
To elucidate allometric scaling in complex systems, we investigated the underlying scaling relationships between typical three-scale indicators for approximately 500,000 Japanese firms; namely, annual sales, number of employees, and number of business partners.  First, new scaling relations including the distributions of fluctuations were discovered by systematically analyzing conditional statistics. Second, we introduced simple probabilistic models that reproduce all these scaling relations, and we derived relations between scaling exponents and the magnitude of fluctuations.
\end{abstract}

\pacs{89.75.Da, 89.75.Fb, 89.65.Gh}

\maketitle

\section{Introduction}
In physiology and anatomy, ``allometric scalings" are empirical power laws among percentiles related to size. For example, the brain mass of mammals scales as the corresponding body mass to the power about $0.7$ \cite{stahl1965organ}.
One of the most famous laws in this field is that between body mass and metabolic rate (i.e., the speed of metabolism), which has the scaling exponent $2/3$ \cite{Kleiber1947}. From the viewpoint of statistical physics, this nontrivial scaling relation is explained by the geometric structure of vessel networks and an assumption regarding minimum energy consumption \cite{West1997}. \par

Recently, allometric scalings have been observed in the real world in various complex systems other than biological systems, and there are many attractive societal applications; for example, economic indices as a function of urban population \cite{Bettencourt2007, Bettencourt2010} energy consumptions vs urban population \cite{horta2010}, surface area of roads vs that of cities \cite{Samaniego2008},
or economic indices vs national populations \cite{Zhang2010}. \par

Fluctuations associated with these scalings in complex systems have also been studied. In these studies, the distribution of growth rates is one of the main topics, and the width of growth rates (e.g., the standard deviation or the interquartile distance) vs system size has been found to follow a power law with a negative exponent \cite{Stanley1996,Labra2007}.
In accordance with this scaling, the conditional distributions of growth rates normalized by the widths or the standard deviations conditioned by the system size collapse onto universal curves, which are independent of system size.
Such conditional distributions of growth rates have been reported for sales of business firms \cite{Stanley1996,fujiwara2004pareto,fu2005growth},  national gross domestic products \cite{fu2005growth}, university research activities \cite{plerou1999similarities}, citations to scientific journals\cite{mendes2006scaling}, the circulation of magazines and newspapers \cite{mendes2005statistical}, religious activities \cite{picoli2008universal}, birds populations \cite{keitt1998dynamics} and the metabolic rates of animals \cite{Labra2007} etc. This characteristic is also commonly observed between the metabolic rates of animals and business firms \cite{Labra2007}. \par
Here, we focus on the statistical properties of business firms and regard each firm as a typical complex system consisting of various elements such as employees, facilities, and money. Firm activity, in the form of financial reports, is rendered numerically observable. The data within typical financial reports contains many quantities relating to firm size, which we can roughly categorize into three families:
\begin{enumerate} 
\item Flow variables; such as annual sales, profit, incomes, or tax payments. 
\item Stock variables; such as the number of employees, number of branches, or number of factories. 
\item Business relations; such as the number of business partners or number of affiliated firms. 
\end{enumerate}
\par
Quite interestingly, one body of statistics based on these quantities is generally approximated by a power law distribution that is typically independent of country and observation year; namely, the universal Zipf law for annual sales or profits \cite{axtell2001zipf, fujiwara2004pareto, okuyama1999zipf}. 
There have been many attempts, typically based on mathematical toy models based on stochastic scale-free dynamics, to clarify why such a power law should hold for a one-body distribution \cite{aoyama2000pareto}. \par 

A few pioneering works exists on allometric scaling of business firms. For example, Fujiwara et.al. reported that employee numbers and incomes scale with the corresponding universal conditional distribution for Japanese business firms (up to intermediate size) \cite{Aoyama2010}, Watanabe et.al have also analyzed these financial scalings by using the production function \cite{Watanabe2011mega} and Saito et.al have showed a scaling relationship between numbers of business partners and annual sales \cite{Saito2007}.  \par
  
In this study we analyze two- and three-body statistics of typical business variables from the three data categories of annual sales, number of employees, and number of business partners. In particular, we focus on the relation between the scalings among the three quantities and the fluctuations associated with them. By analyzing data from about $500,000$ Japanese firms, we find in Sec. 2 that some pairs of these quantities follow power laws. 
In addition, we show that the distribution functions for different parameters converge to a unique scaling function through these scaling relations of conditional medians. In the same section, we also find, for three-body relations, scalings of the conditional median of sales and employees as a function of the other two variables. In Sec. 3, we introduce simple stochastic models that reproduce the all empirical scalings and discuss the relations between these scalings and fluctuations. Finally, we conclude with a discussion in Sec. 4.

\section{Data analysis}
\label{sec:observation}
The data set was provided by the governmental research institute RIETI (Research Institute of Economy, Trade and Industry) and was based on data collected by Tokyo Shoko Research, Ltd. (TSR) for 2005. It contains approximately one million firms covering practically all active firms in Japan. For each firm, the data set contains various flow variables, stock variables, and a list of business partners categorized into suppliers and customers \cite{Ohnishi2009}.
From this list, we count the total number of business partners, by superposing all business interactions. 
We focus on the three basic scale indicators of firms from the three categories: sales $s$, number of employees $l$, and number of business partners, which we call the degree $k$. 
We neglect those firms for which the three data are not available, thus that the number of firms we analyze is 529,291. \par    

\subsection{Correlations between two variables}
In general, all information regarding three-body statistics for stochastic variables $\{X,Y,Z\}$
 is contained in the three-body probability density function (PDF), $P(X,Y,Z)$. 
To clarify the structure of this function, using the definition of the conditional probability, 
we decompose it into the three density functions as $P(X,Y,Z)=P(X|Y,Z)P(Y|Z)P(Z)$, where we denotes the conditional probability density of $Y$ for given value of $Z$ by $P(Y|Z)$, and where $P(X |Y, Z)$ is the conditional probability density of $X$ for simultaneously given values of $Y$ and $Z$. 
We pay attention to the properties of these conditional probability densities.
Firstly, we are going to observe the probability densities conditioned by one variable, $P(Y|Z)$, and then the probability densities conditioned by two variables $P(X|Y,Z)$. \par

\begin{figure}
\begin{minipage}{1\hsize}
\centering
\includegraphics[width=6.3cm]{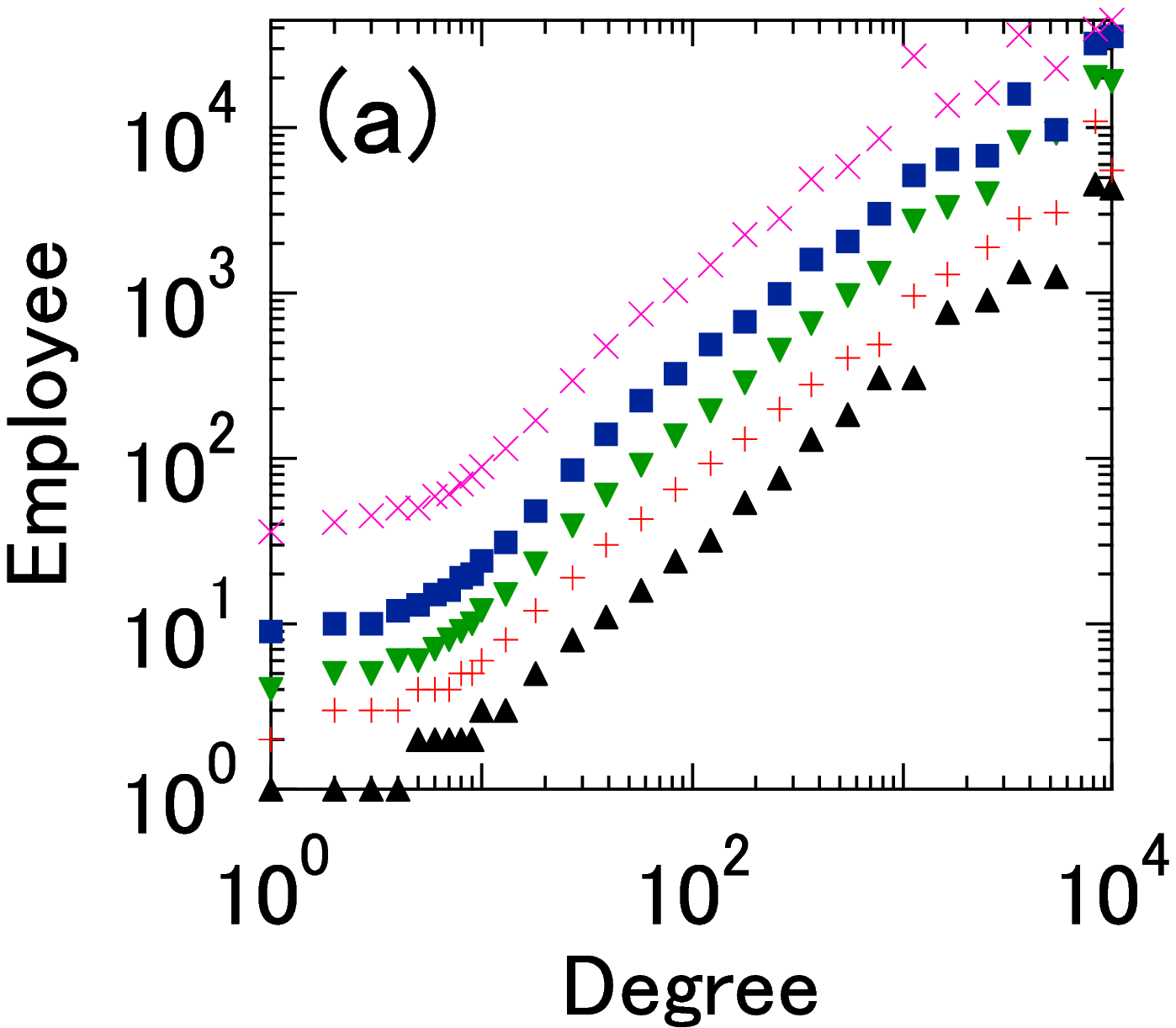}
\end{minipage}
\begin{minipage}{1\hsize}
\centering
\includegraphics[width=6.3cm]{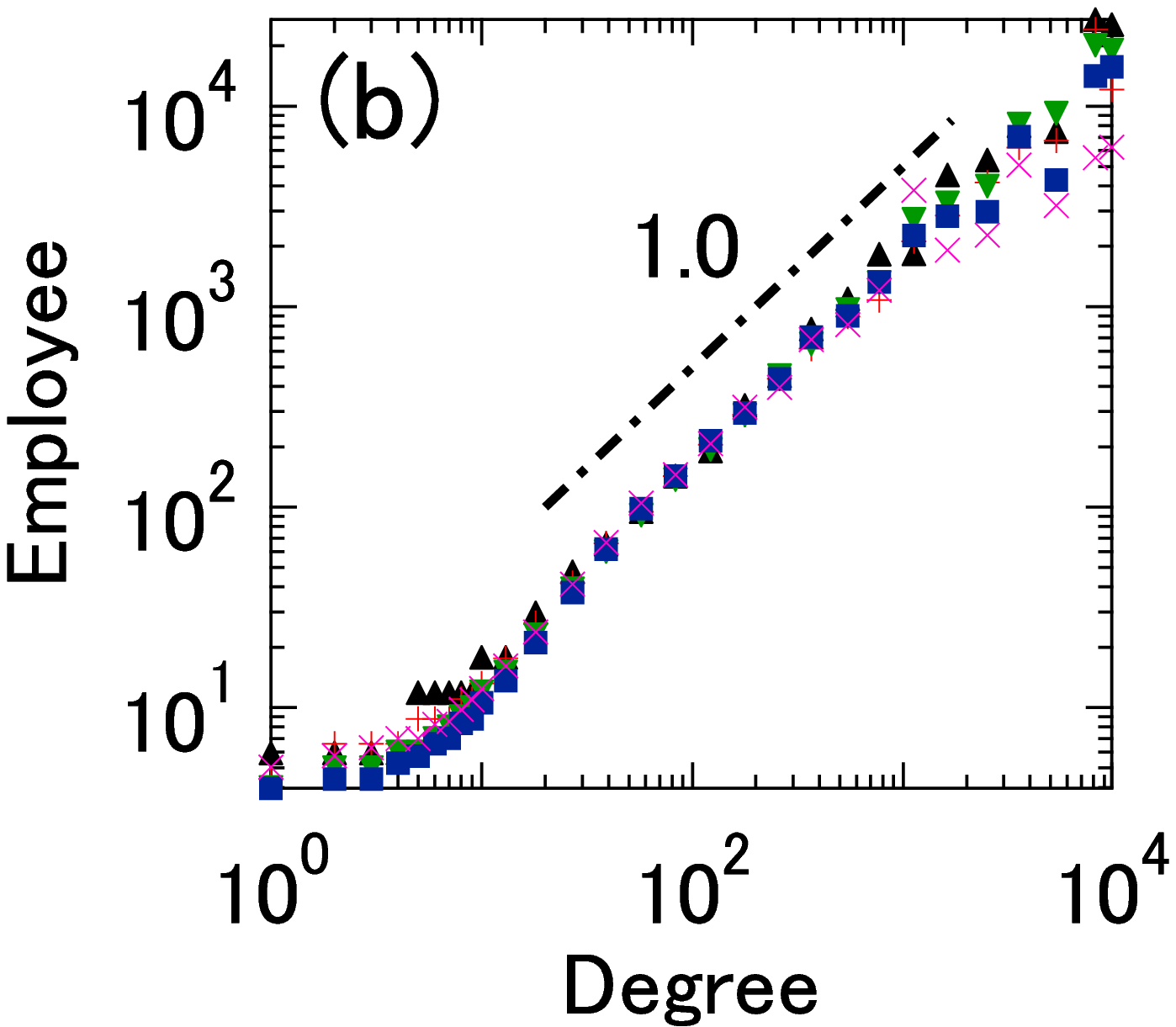}
\end{minipage}
\begin{minipage}{1\hsize}
\centering
\includegraphics[width=6.3cm]{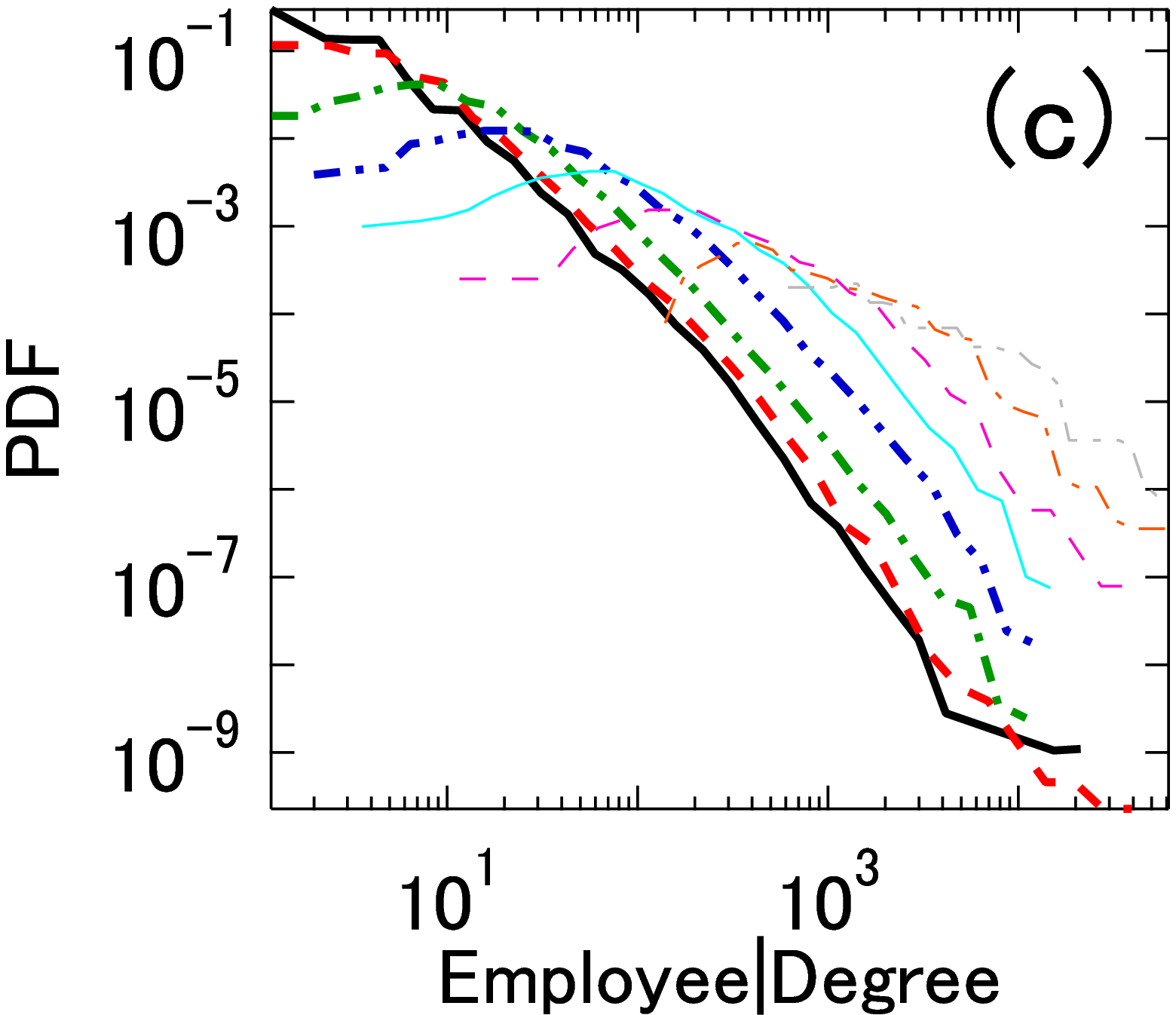}
\end{minipage}
\begin{minipage}{1\hsize}
\centering
\includegraphics[width=6.3cm]{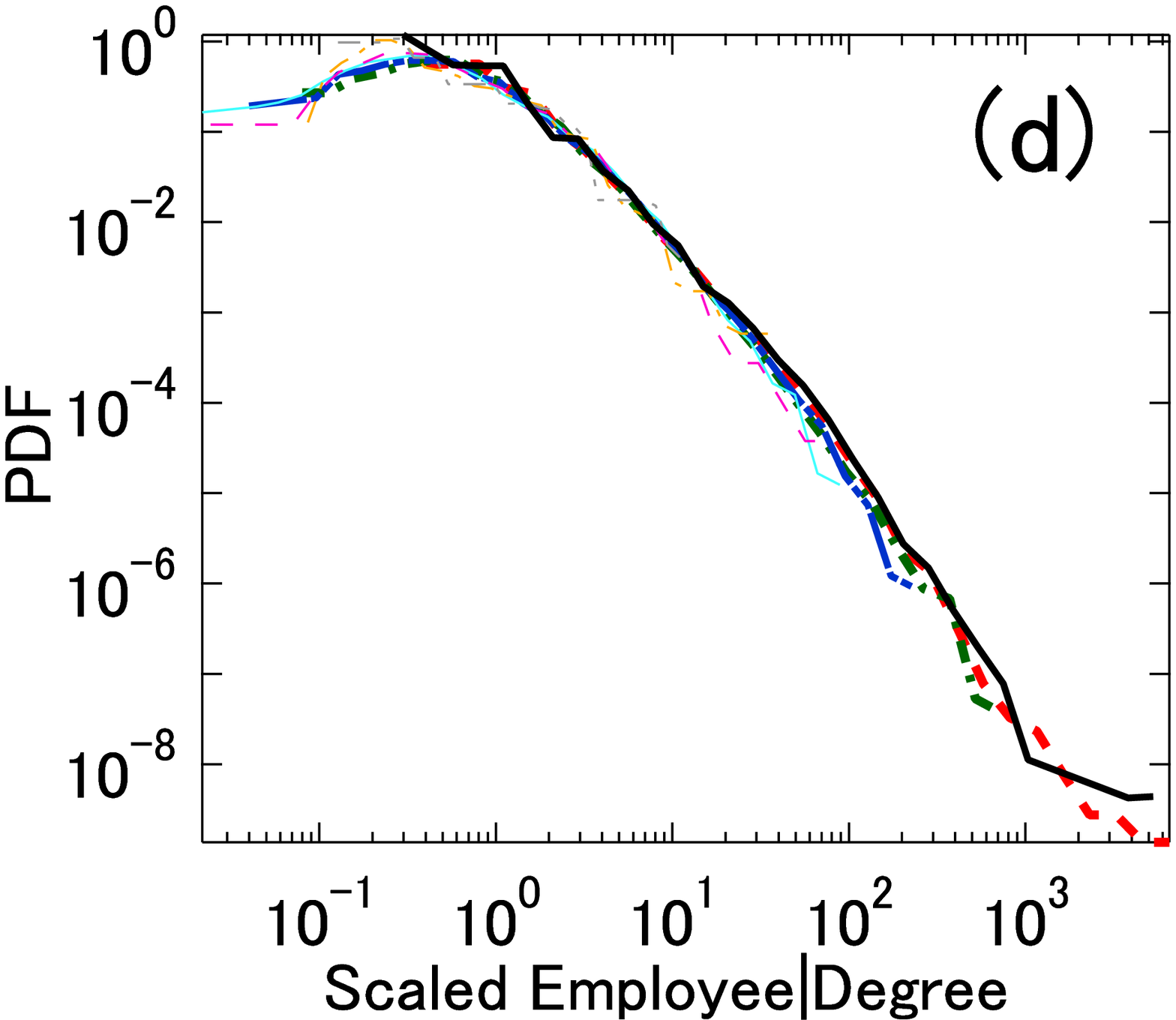}
\end{minipage}
\caption{Scaling relations for number of employees conditioned by degrees. 
(a)Conditional percentile of the number of employees $l$ given degree $k$. The data shown are 5th percentile (black triangles), 25th percentile (red plus signs), 50th percentile (green nablas) ,75th percentile (blue squares), and 95th percentile (purple crosses). 
 (b)Corresponding percentiles obtained by shifting plots in panel (a), along the vertical axis 
so that they overlap with the median plots at $k=100$.
  The black dashed-dotted line shows the slope of $k^{1.0}$.
  (c)Conditional PDF of number of employees $l$ for given degree $k$, $P(l|k)$, where the conditional parameter $k$ was evenly divided in logarithmic space into eight boxes. (d)Conditional PDF of the normalized number of employees $\bar{l_k}=l/<l|k>_{0.5}$ given degree $k$. }
\label{degree_employee}
\end{figure}

We begin by analyzing the two-body relations between the number of employees $l$ and degree $k$.
Fig. \ref{degree_employee}(a) shows the log-log plot of the number of employees as a function of degree $k$. 
We find that all such plots have similar forms for the 5th, 25th, 50th (equivalent to the median), 75th, and 95th percentiles of the number of employees $l$ for a given degree $k$.  
In Fig. \ref{degree_employee}(b), we shift these plots along the vertical axis so that they all lie on the median plot at $k=100$.
All these conditional percentile curves essentially coincide with each other. 
In particular, for $k \geq 30$, this relation can be described by the following scaling relation:
\begin{equation}
<l|k>_{q}=B^{(l|k)}_q \cdot k^{\gamma_{l|k}} \quad (q=0.05,0.25,0.5,0.75,0.95), \label{l_k_kika}
\end{equation}
where $\gamma_{l|k}=1.0$, $<l|k>_{q}$ is the $100q$ conditional percentile of $l$ given $k$ and $B^{(l|k)}_q$is a proportional constant for percentile $100q$.
The values of $B^{(l|k)}_q$ are estimated to be 0.3 for the 5th percentile, 0.7 for the 25th percentile, 1.6 for the 50th percentile, 4.0 for the 75th percentile and 12 for the 95th percentile. 
$B^{(l|k)}_q$ can be interpreted as the number of employees per business partner.
 We find the typical value at the median is $1.6$. 
According to these percentile scaling relations, the PDF of $l$ for a given value of $k$, $P(l|k)$, 
is 
\begin{equation}
P(l|k)=\frac{1}{f_1(k)} \cdot \Psi_1(\frac{l}{f_1(k)}), \label{phi1}
\end{equation}
where $f_1(k)=<l|k>_{0.5}$ is the scaling function and $\Psi_1(\cdot)$ is the PDF of the normalized quantity, $\bar{l_k} \equiv l/f_1(k)$, which does not depend on $k$. 
Noted that, because the lower limit of the number of employees is 1, the PDF has a cut off for small $k$. 
In Fig. \ref{degree_employee}(c), 
we plot the conditional PDF $P(l|k)$ for several values of $k$, which shifts right with increasing degree $k$. 
In Fig. \ref{degree_employee}(d), we see that plot of $\Psi_1(\bar{l_k})$ actually does not depend on the value of $k$. 
\par
\begin{figure}
\begin{minipage}{1\hsize}
\includegraphics[width=6.3cm]{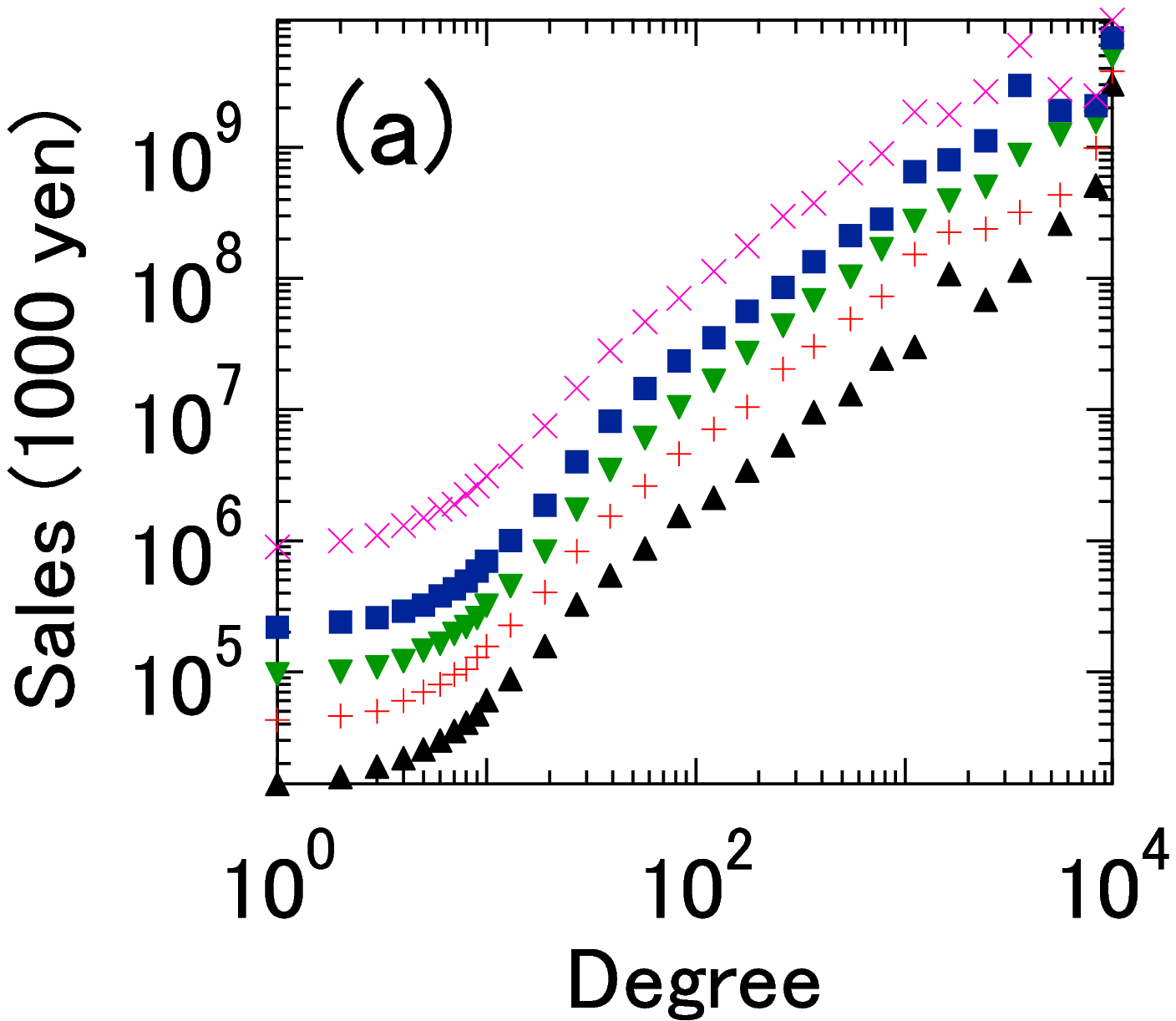}
\end{minipage}
\begin{minipage}{1\hsize}
\includegraphics[width=6.3cm]{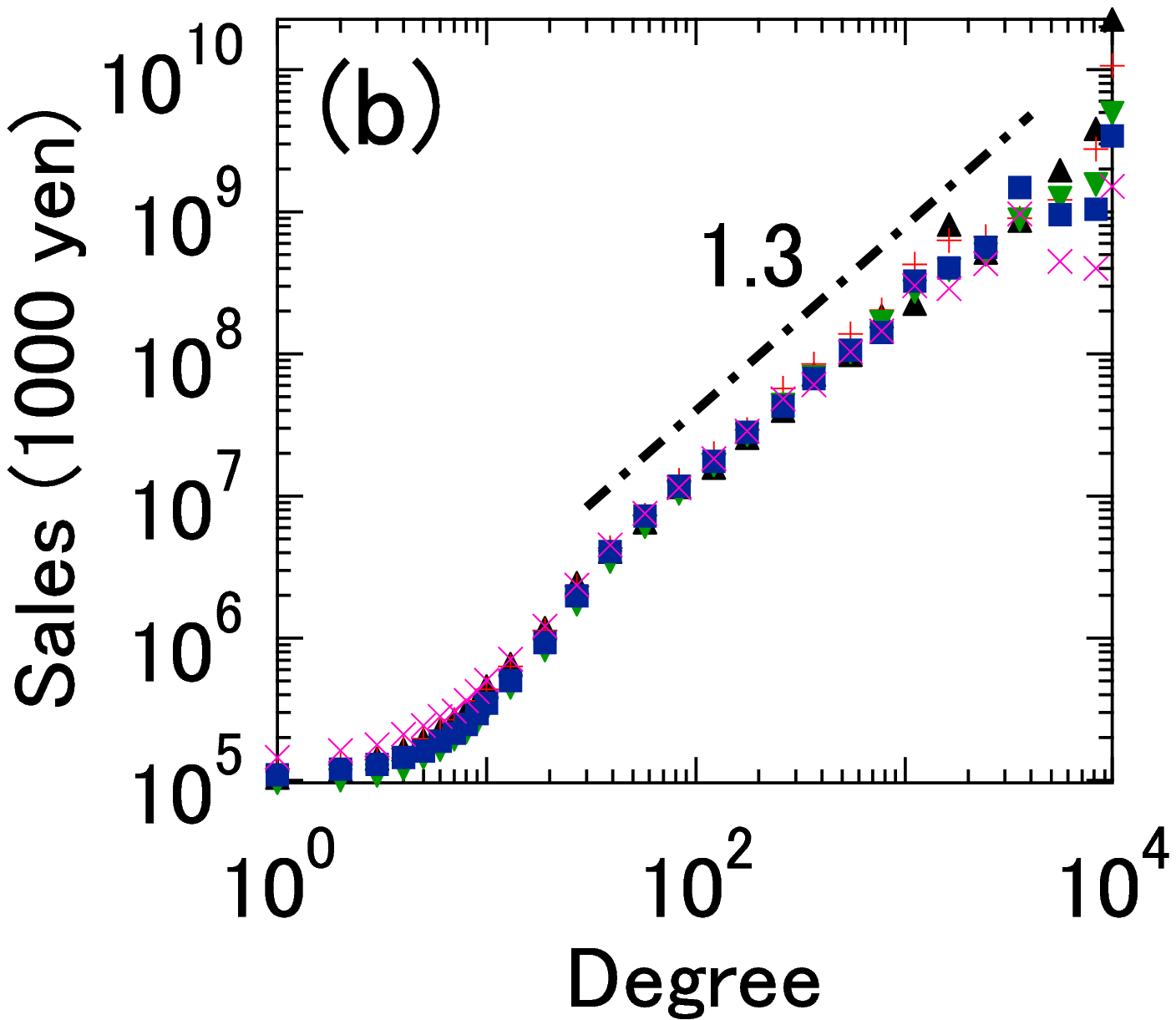}
\end{minipage}
\begin{minipage}{1\hsize}
\includegraphics[width=6.3cm]{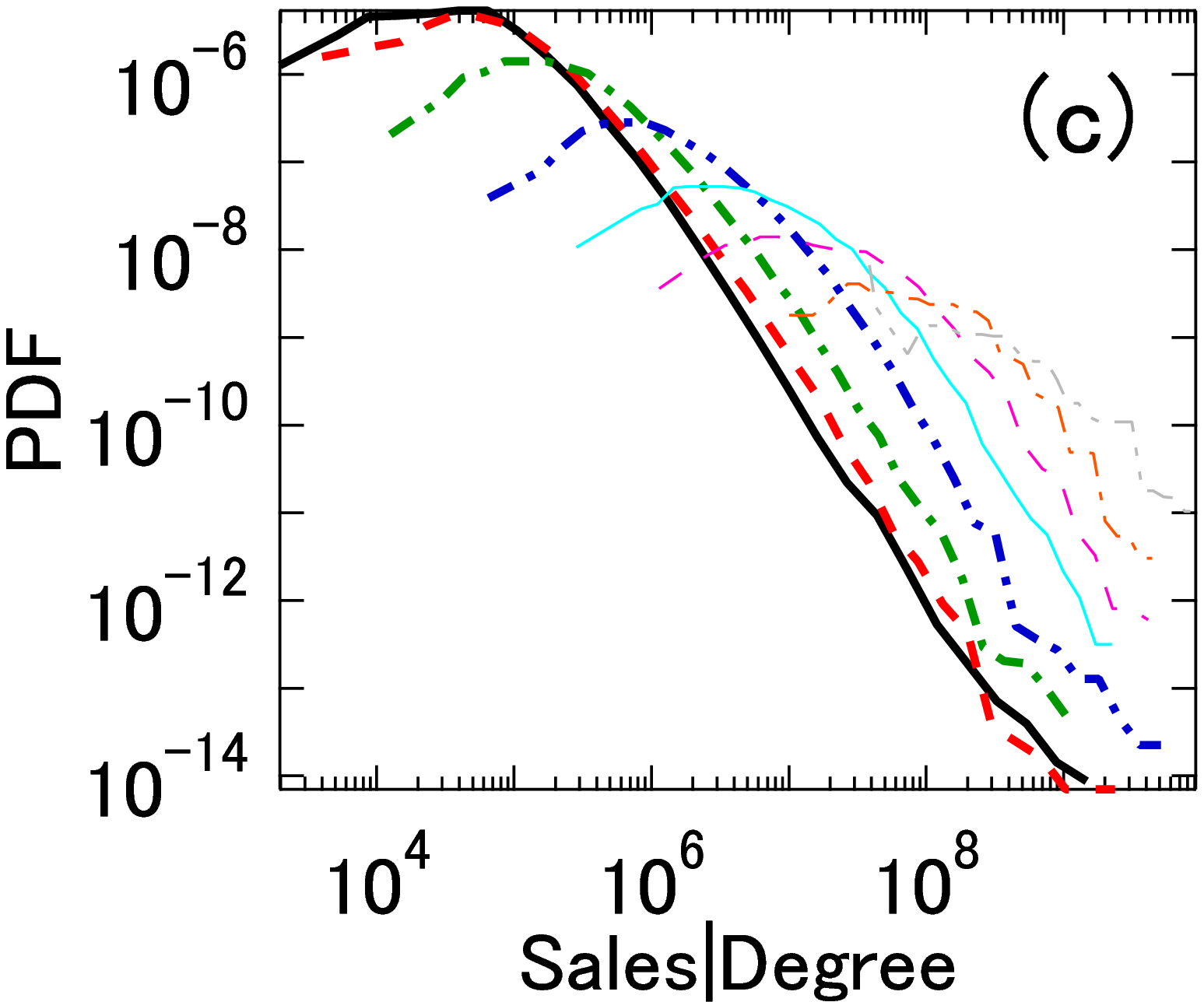}
\end{minipage}
\begin{minipage}{1\hsize}
\includegraphics[width=6.3cm]{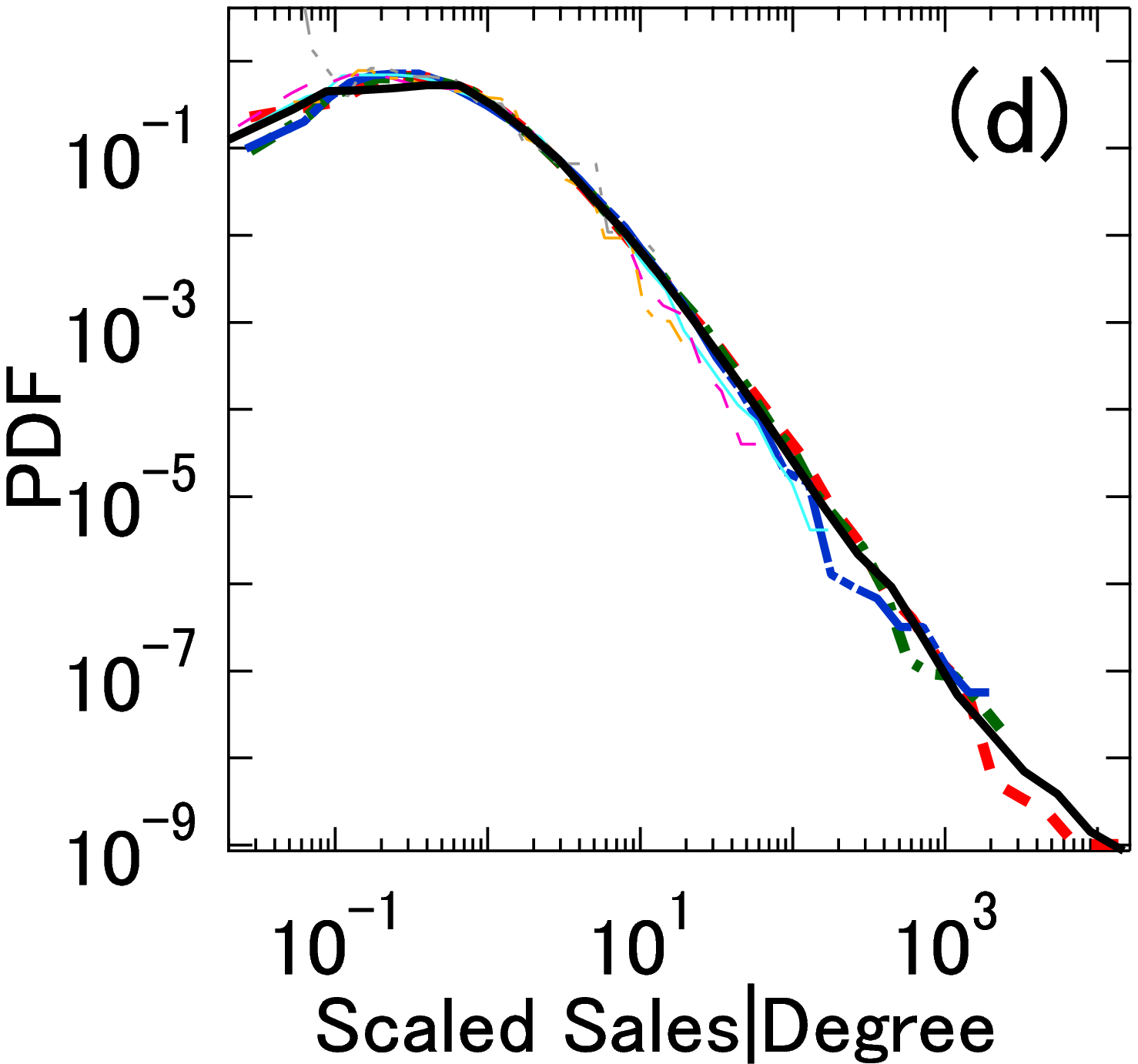}
\end{minipage}
\caption{
Scaling relations of sales conditioned by degrees. 
(a)Conditional percentile of sales $s$ given degree $k$. 
The data for the 5th percentile (black triangles), 25th percentile (red pluses signs),
 50th percentile (green nablas) ,75th percentile (blue squares),
 and 95th percentile (purple crosses). 
(b)Corresponding percentiles obtained by shifting plots in panel (a),  along the vertical axis 
so that they overlap with the median plot at $k=100$.
 The black dashed-dotted line shows a slope of $k^{1.3}$.  
 (c)Conditional PDF of sales $s$ for given degree $k$, $P(s|k)$, where the conditional parameter $k$ was evenly divided into eight boxes in logarithmic space. (d)Conditional PDF of normalized sales $\bar{s_k}=s/<s|k>_{0.5}$ given degree $k$.
}
\label{degree_sales}
\end{figure}

We apply a parallel analysis for relations between sales $s$ and the degree $k$.
Thus, Figs. \ref{degree_sales}(a) and (b), we can confirm that, for all range of $k$, 
all of these conditional percentile curves essentially coincide with each other after shifting them along the vertical axis.
For $k$ ranging from $30$ to $1000$, the following nontrivial scaling relation holds 
for the conditional percentile values $<s|k>_q$:
\begin{equation}
<s|k>_{q} = B^{(s|k)}_q \cdot k^{\gamma_{s|k}} \quad (q=0.05,0.25,0.5,0.75,0.95), \label{s_k_kika}
\end{equation}
where $\gamma_{s|k}=1.3$, $B^{(s|k)}_{0.05}=3.8 \cdot 10^6$ (yen), $B^{(s|k)}_{0.25}=14\cdot10^6$ (yen), $B^{(s|k)}_{0.5}=32 \cdot 10^6$ (yen), $B^{(s|k)}_{0.75}=60 \cdot 10^6$ (yen) and 
 $B^{(s|k)}_{0.95}=190 \cdot 10^6$ (yen).
Note that this scaling exponent value, $\gamma_{s|k}=1.3$, differs significantly from that for the employees, $\gamma_{l|k}=1.0$.
This result implies that the mean of ``sales per degree" increases with increasing number of business partners. 
The conditional PDFs of sales for different $k$, 
$P(s|k)$ are plotted in Fig. \ref{degree_sales}(c).
This function is expected to be expressed by a scaling function as
\begin{equation}
P(s|k)=\frac{1}{f_2(k)} \cdot \Psi_2(\frac{s}{f_2(k)}) \label{phi2}
\end{equation}
where $f_2(k)=<s|k>_{0.5}$ is the scaling between $s$ and $k$ at the median point. 
The PDF of sales normalized by using this scaling, $\bar{s_k} \equiv s/f_2(k)$, 
$\Psi_2(\cdot)$ is plotted in Fig. \ref{degree_sales}(d), and we confirm that $\Psi_2$ is 
independent of degree $k$. \par

\begin{figure}
\begin{minipage}{1\hsize}
\centering
\includegraphics[width=6.3cm]{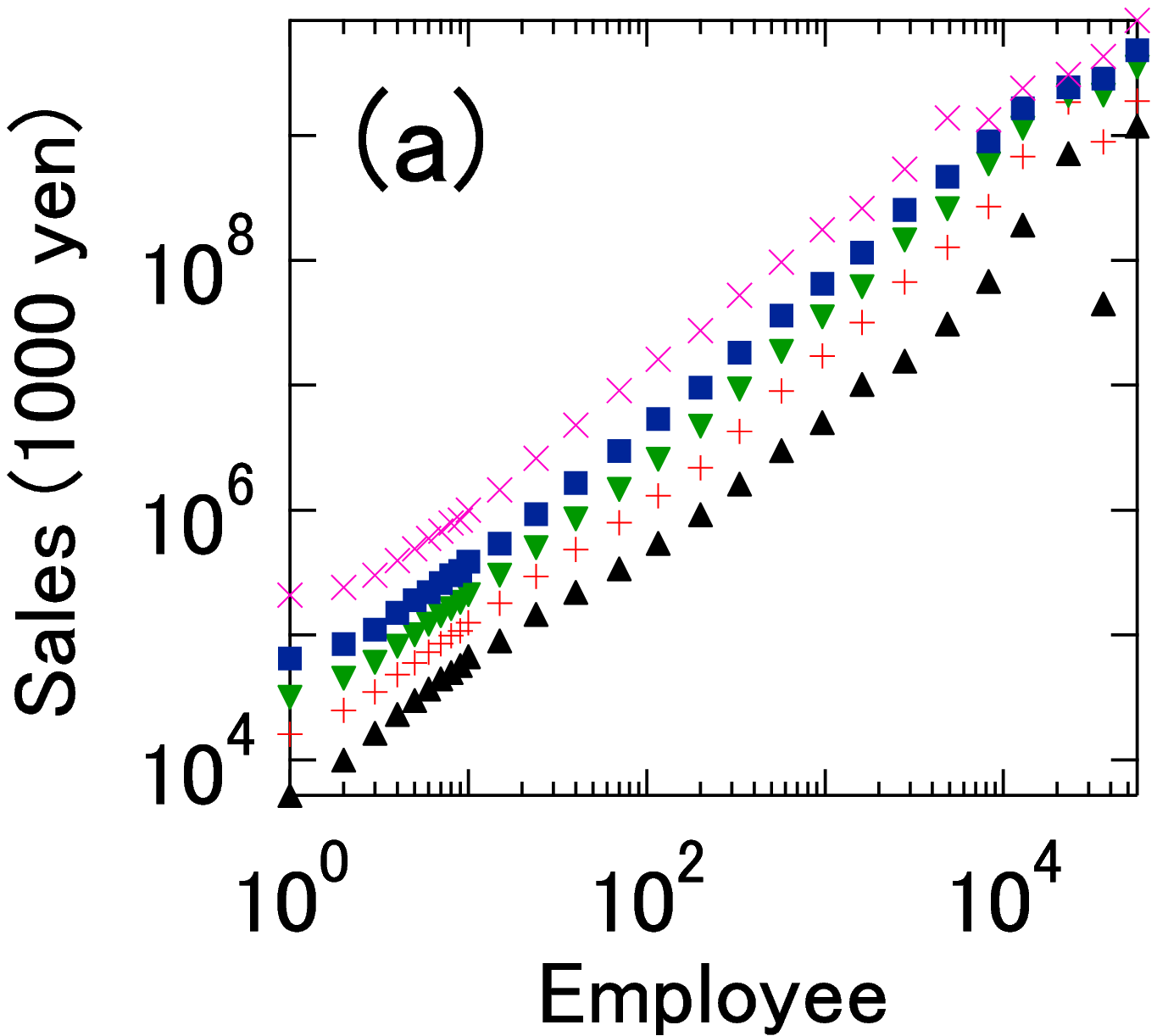}
\end{minipage}
\begin{minipage}{1\hsize}
\centering
\includegraphics[width=6.3cm]{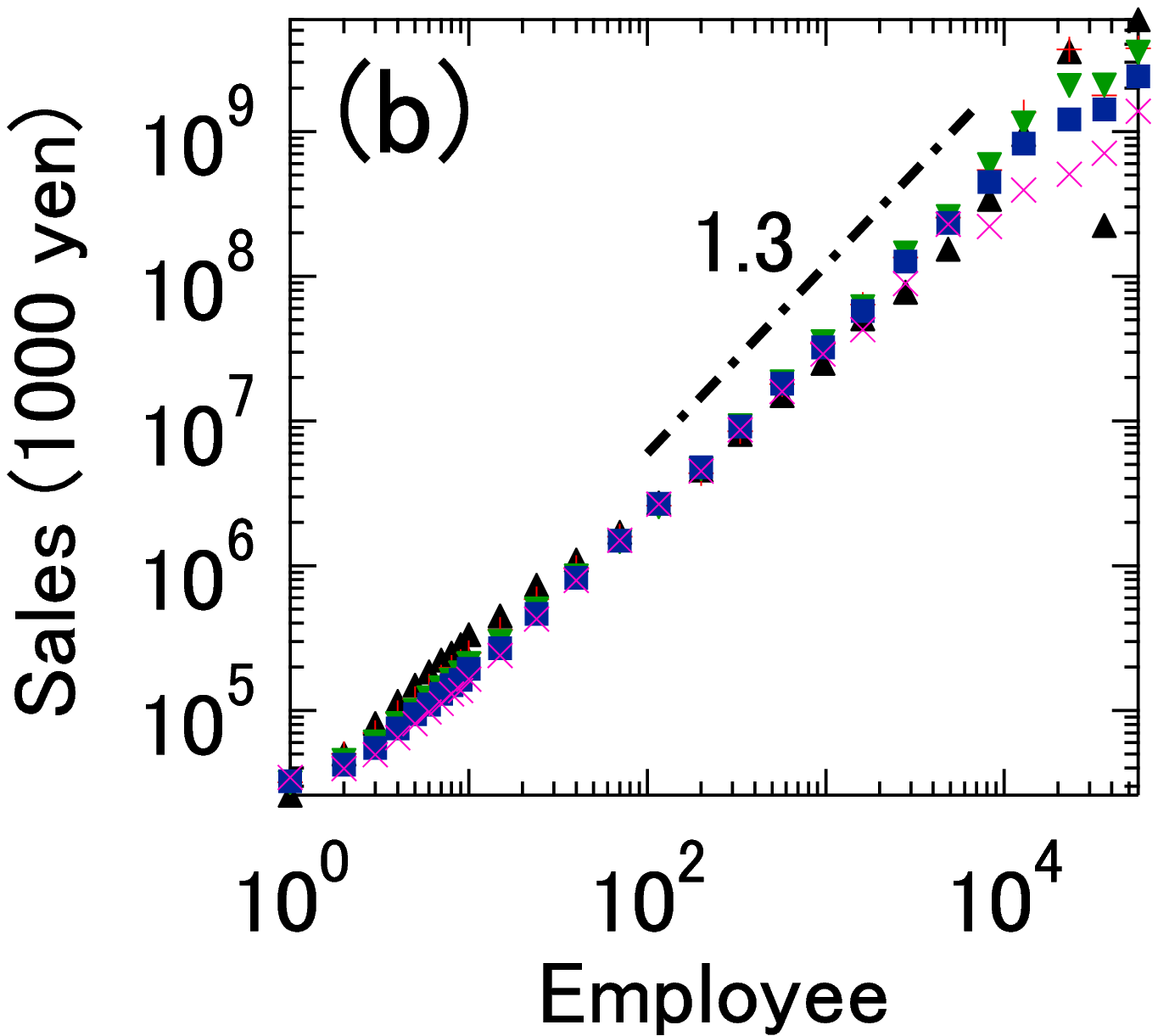}
\end{minipage}
\begin{minipage}{1\hsize}
\centering
\includegraphics[width=6.3cm]{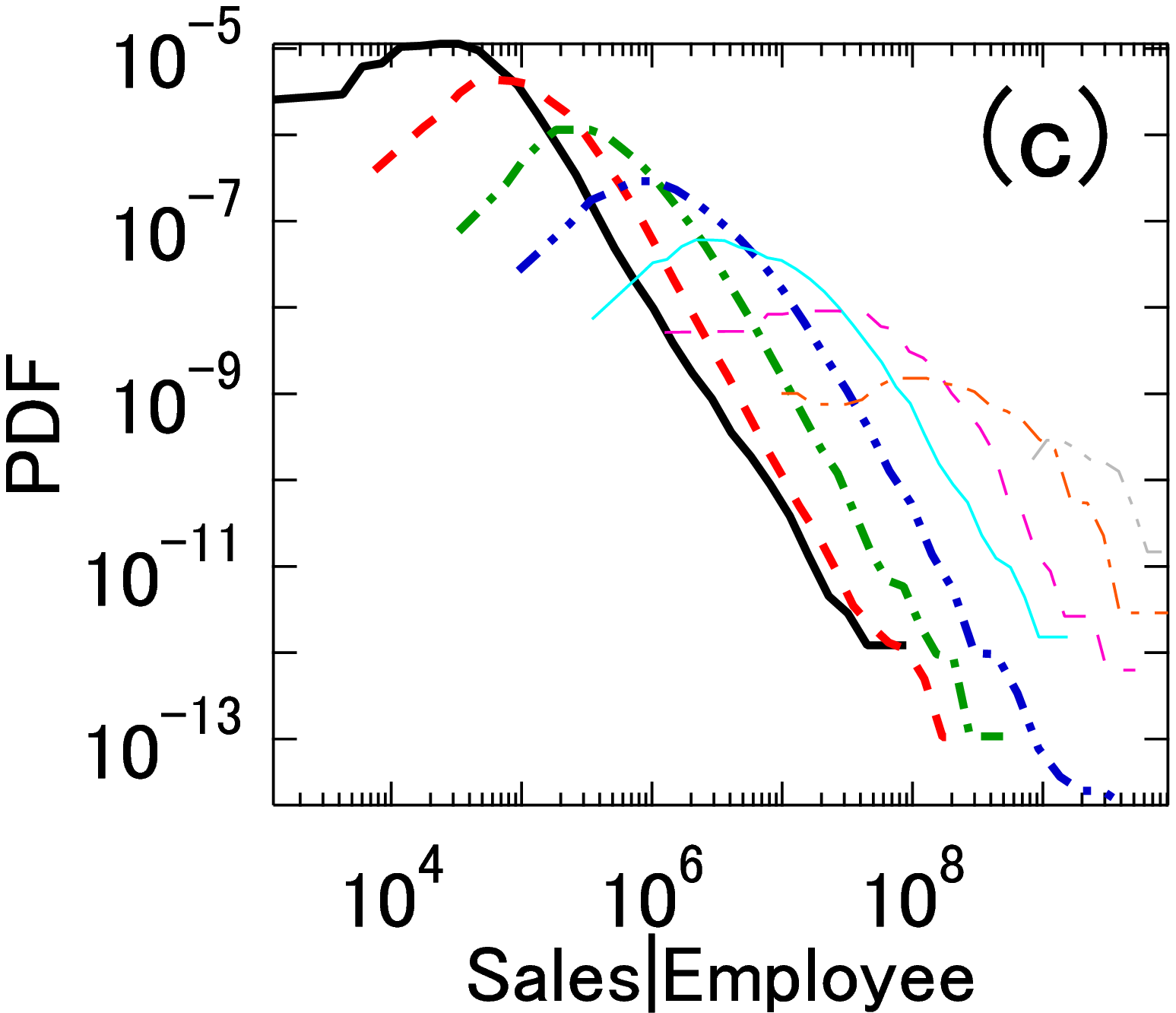}
\end{minipage}
\begin{minipage}{1\hsize}
\centering
\includegraphics[width=6.3cm]{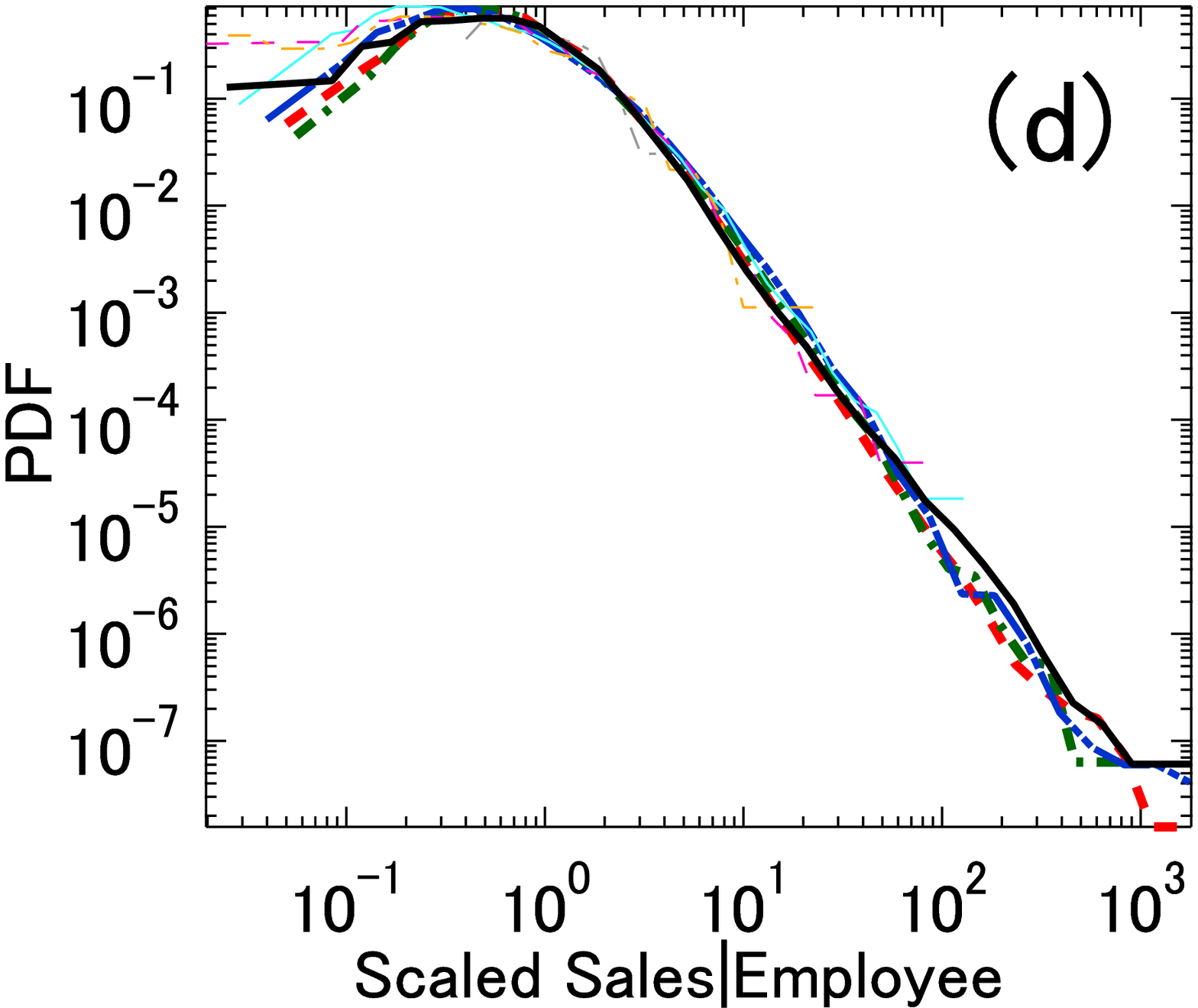}
\end{minipage}
\caption{Scaling relations of sales conditioned by number of employees. 
(a)Conditional percentile of sales $s$ for a given number of employees $l$. 
The data shown are 5th percentile (black triangles), 25th percentile (red pluses signs),
 50th percentile (green nablas), 75th percentile (blue squares),
 and 95th percentile (purple crosses). 
(b)Corresponding percentiles obtained by shifting these plots so that they along the vertical axis to overlap with the median curve at $l=100$.
 The black dashed-dotted line has a slope of $l^{1.3}$.  
 (c)Conditional PDF of sales $s$ for a given number of employees $l$, $P(s|l)$, where the conditional parameter $l$ is evenly divided  in the logarithmic space into eight boxes. 
 (d)Conditional PDF of normalized sales $\bar{s_l}=s/<s|l>_{0.5}$ for a given number of employees $l$. }
\label{employee_sales}
\end{figure}

Finally, we investigate the relation between sales and number of employees by a parallel analysis just like that of the other pairs.
As shown in Figs. \ref{employee_sales}(a) and 3(b), we obtain the 
following scaling relation between sales and number of employees:
\begin{equation}
<s|l>_{q} = B^{(s|l)}_q \cdot l^{\gamma_{s|l}} \quad (q=0.05,0.25,0.5,0.75,0.95), \label{s_l_kika}
\end{equation}
where $<s|l>_q$ denotes the percentile $100q$ of sales given by employee numbers $l$, $\gamma_{s|l}=1.3$, $B^{(s|l)}_{0.05}=5.5 \cdot 10^5$ (yen), $B^{(s|l)}_{0.25}=22 \cdot 10^5$ (yen), 
$B^{(s|l)}_{0.5}=47 \cdot 10^5$ (yen), $B^{(s|l)}_{0.75}=80 \cdot 10^5$ (yen) and $B^{(s|l)}_{0.95}=19 \cdot 10^6$ (yen).  
The conditional PDF of sales, $P(s|l)$, 
is plotted in Fig. \ref{employee_sales}(c) and the corresponding PDF of the normalized variable, 
$\Psi_3(s/f_3(l))$ is plotted in Fig. \ref{employee_sales}(d).  The normalized variable $\Psi_3(s/f_3(l))$ 
is defined by,   
\begin{equation}
P(s|l)=\frac{1}{f_3(l)} \cdot \Psi_3(\frac{s}{f_3(l)}) \label{phi3},
\end{equation}
where $f_3(l)=<s|l>_{0.5}$.  
The results shown in Fig. \ref{employee_sales}(d) demonstrate
that all conditional PDFs collapse into a single function as expected.
This scaling relation agrees with Eqs (\ref{s_k_kika}) and Eqs (\ref{phi2})

\begin{figure}
\begin{minipage}{1\hsize}
\centering 
\includegraphics[width=7.3cm]{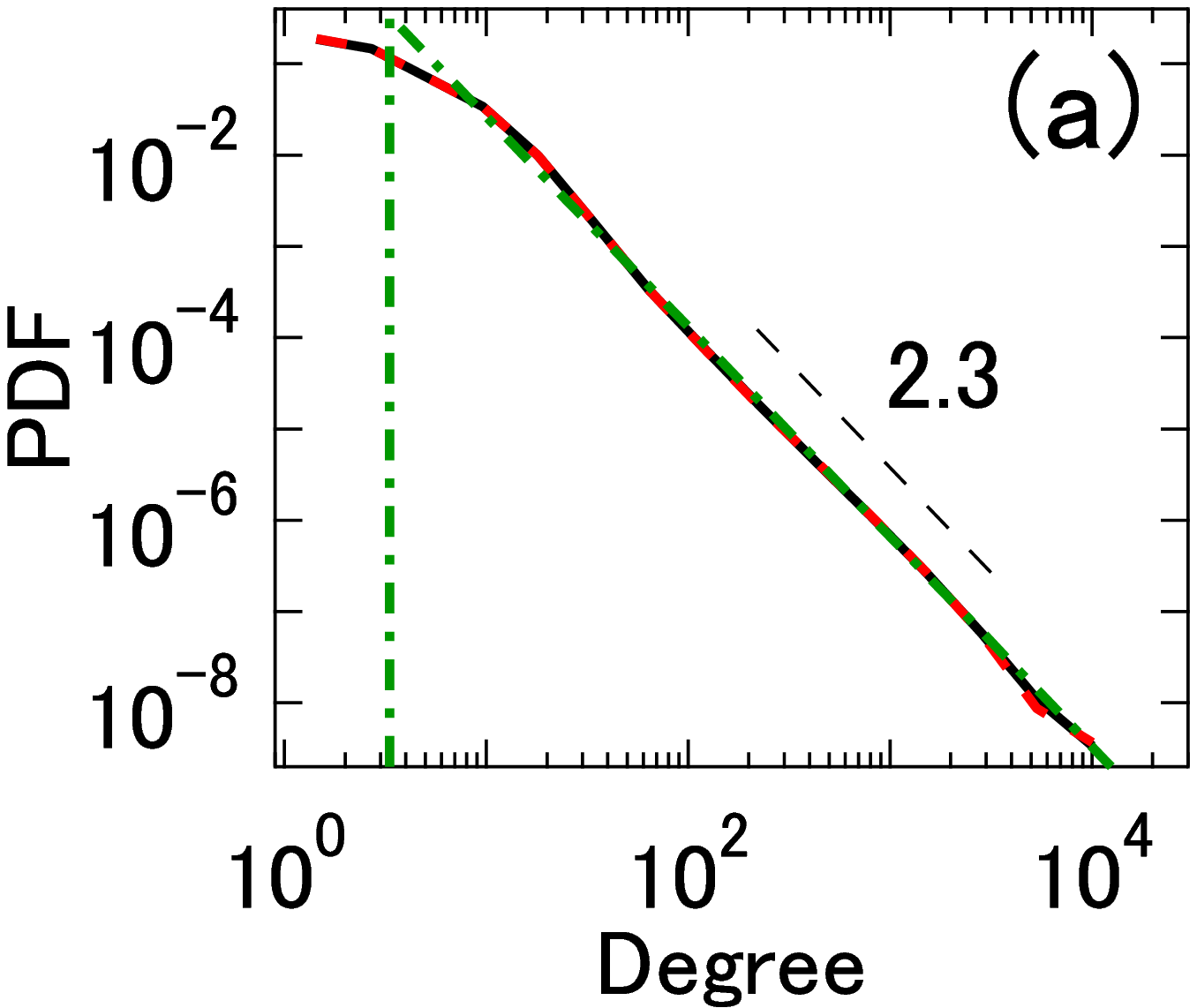}
\end{minipage}
\begin{minipage}{1\hsize}
\centering
\includegraphics[width=7.3cm]{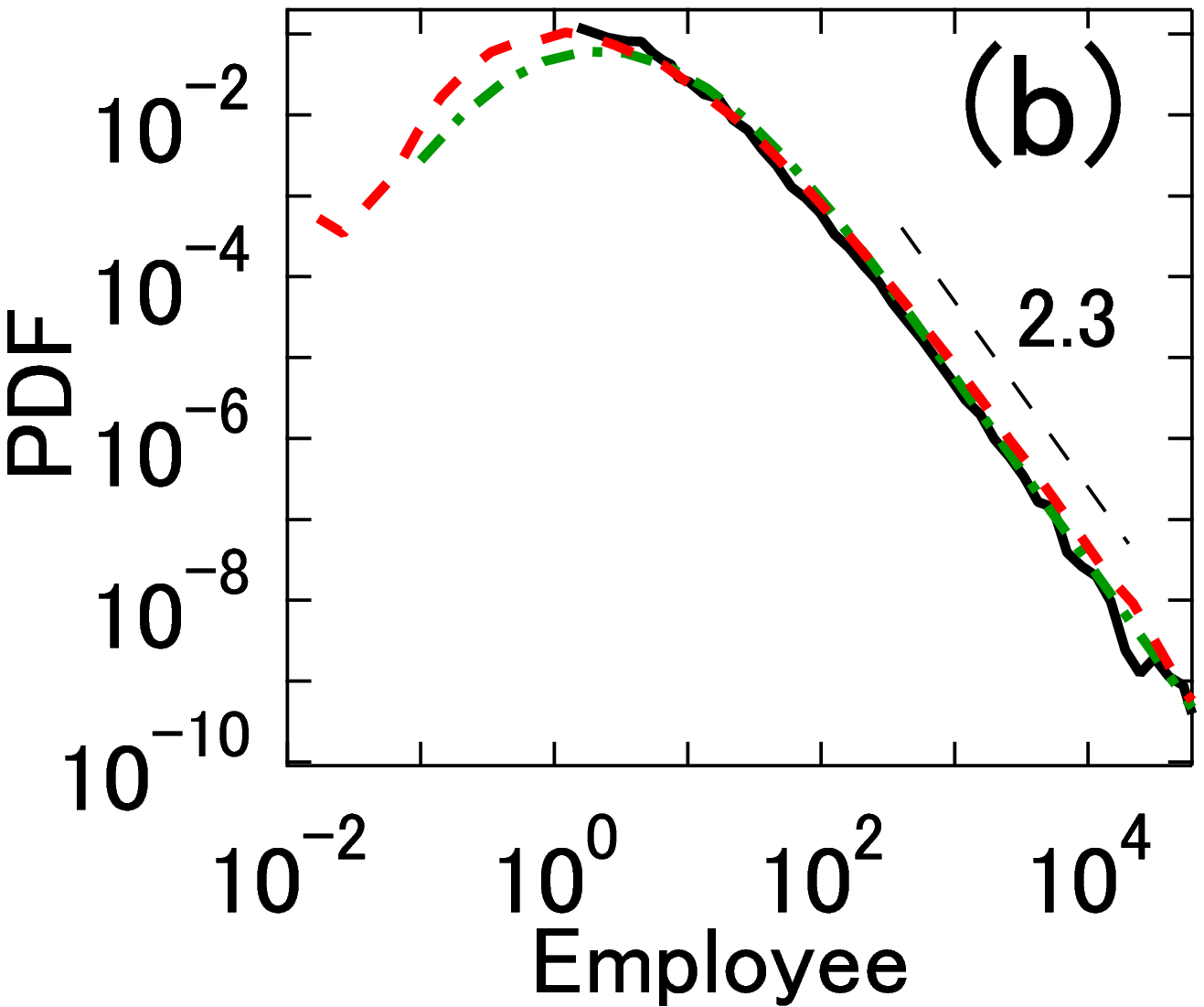} 
\end{minipage}
\begin{minipage}{1\hsize}
\centering
\includegraphics[width=7.3cm]{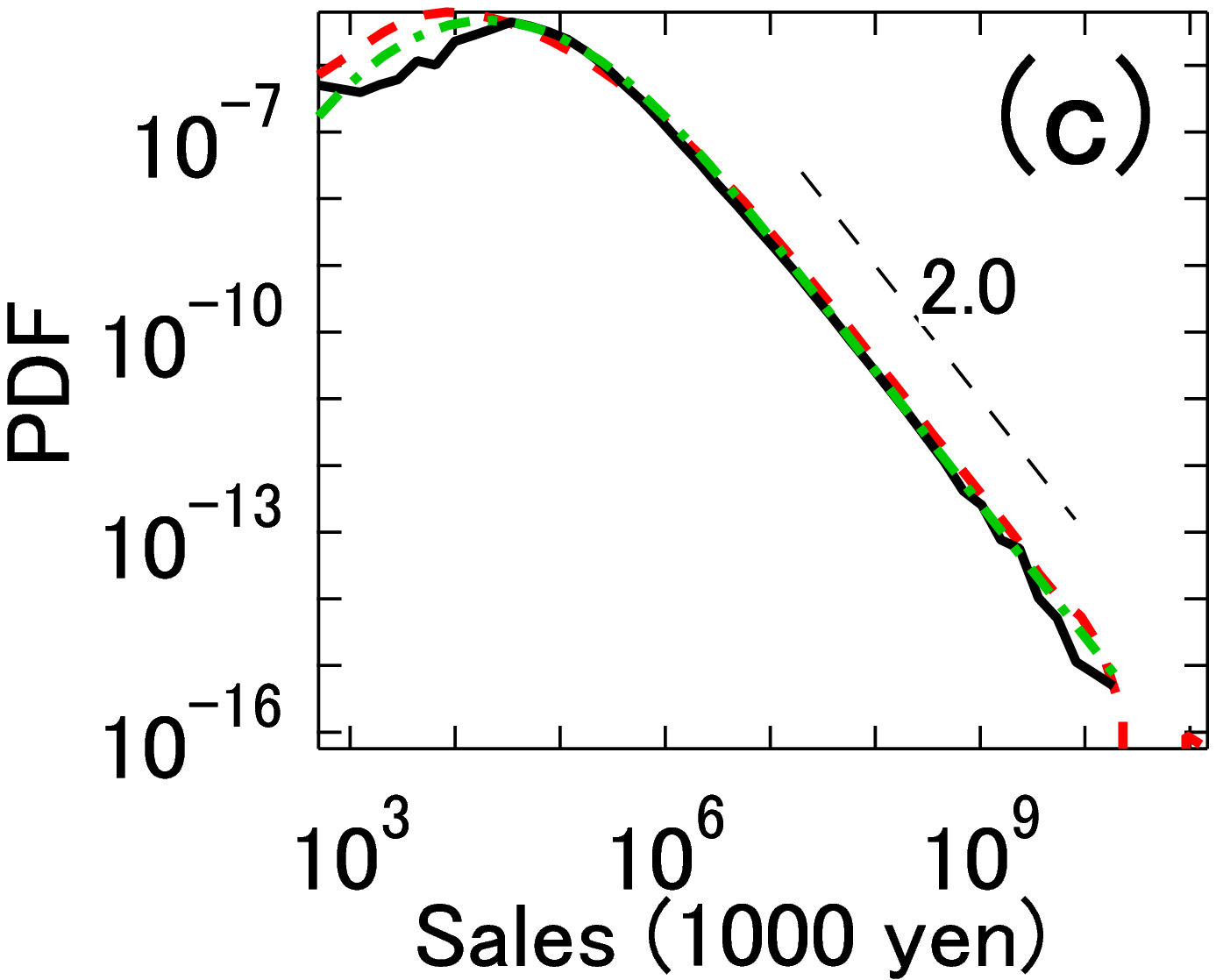}  
\end{minipage}
\caption{
PDFs of degree $k$, number of employees $l$, and sales $s$ for empirical data (black solid lines), shuffled model (red dashed line), and lognormal distribution model (green dash-dotted line). (a) PDFs of degree $k$. The black dashed line shows $k^{2.3}$. 
(b) PDFs of employee $l$. The black dashed line shows $l^{2.3}$. 
(c) PDFs of sales $s$. The black dashed support line shows $s^{2.0}$. 
This figure confirms that sales $s$ obey Zipf's law.
}
\label{sed}
\end{figure}

Integrating over the conditioned variables we have the PDF of a single body variable from the conditioned PDF.
For each variable, $k$, $l$ and $s$, the PDF is plotted in Fig. \ref{sed} (black solid lines) on a loglog scale. 
We have the following power laws: 
\begin{equation}
P(k) \propto k^{-\zeta_k-1};\;P(l) \propto l^{-\zeta_l-1};\; P(s) \propto s^{-\zeta_s-1} \label{power},
\end{equation}
where $\zeta_k=1.3$, $\zeta_l=1.3$ and $\zeta_s=1.0$.
These exponents are directly related to the scaling exponents, as shown below. \par
Assuming that $X$ obeys the following power-law distribution with the PDF: 
\begin{equation}
p_{X}(X) \propto {X}^{-\zeta_{X}-1}
\end{equation}
and also assuming that $X$ and $Y$ satisfy the allometric scaling relation
\begin{equation}
Y \propto X^{\gamma_{Y|X}}, 
\end{equation}
where $\gamma_{Y|X}$ is the scaling exponent, then by a simple variable transformation,
the PDF of $Y$ is given as
\begin{equation}
p_{Y}(Y) \propto p_{X}(X)\cdot \left| \frac{dX}{dY} \right| \propto Y^{-\zeta_{X}/\gamma_{Y|X}-1}.
\end{equation}
Thus, we get the following relation between the power law indices:
\begin{equation}
\gamma_{Y|X}=\zeta_{X}/\zeta_{Y} \label{sisuu}.
\end{equation}
This relation is confirmed in our data analysis, $\gamma_{l|k}=\zeta_{k}/\zeta_{l}=1.0$
, $\gamma_{s|k}=\zeta_{k}/\zeta_{s}=1.3$ and  $\gamma_{l|s}=\zeta_{l}/\zeta_{s}=1.3$.

\begin{figure}
\begin{minipage}{1\hsize}
\includegraphics[width=6.5cm]{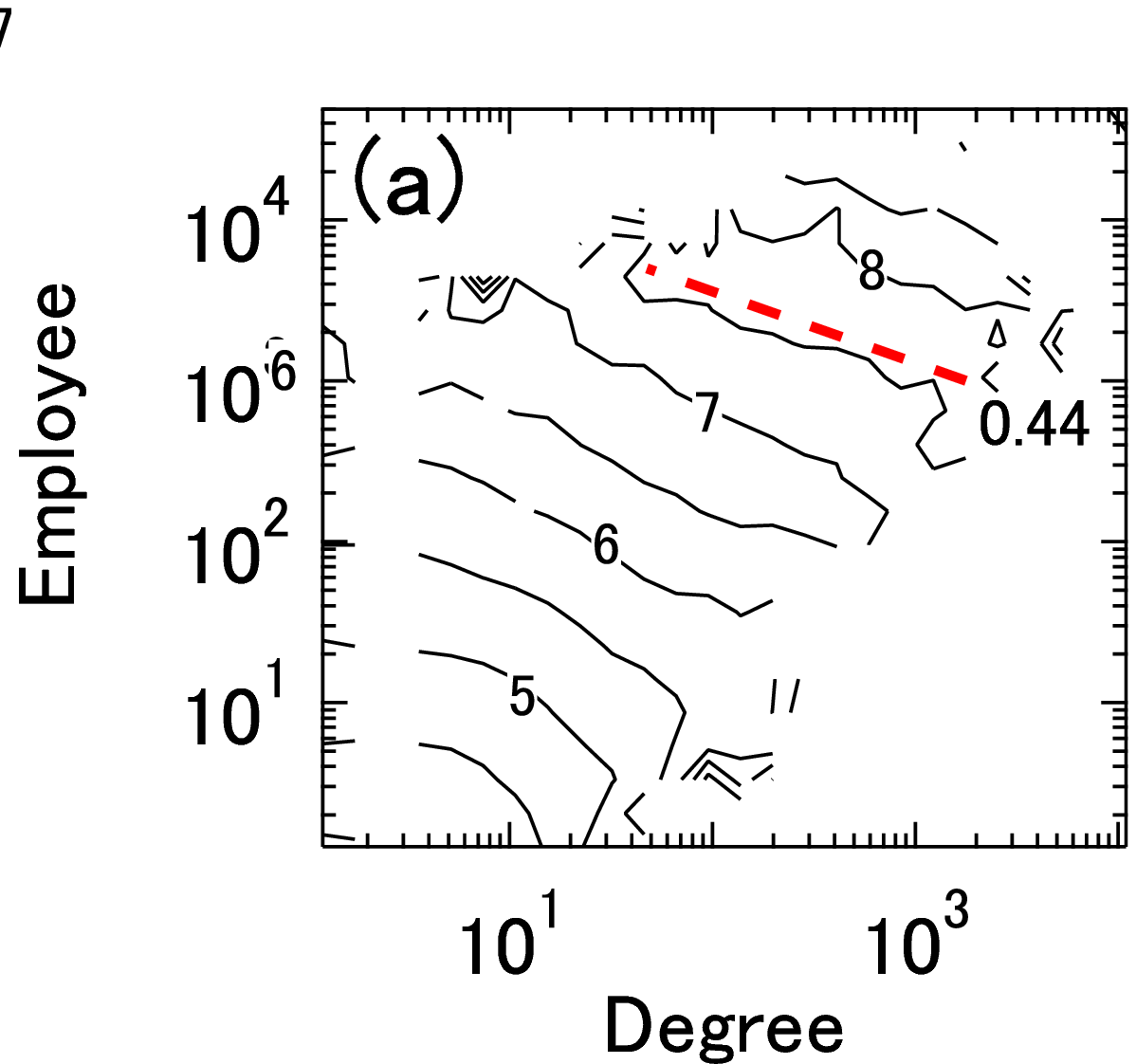}
\end{minipage}
\begin{minipage}{1\hsize}
\includegraphics[width=6.5cm]{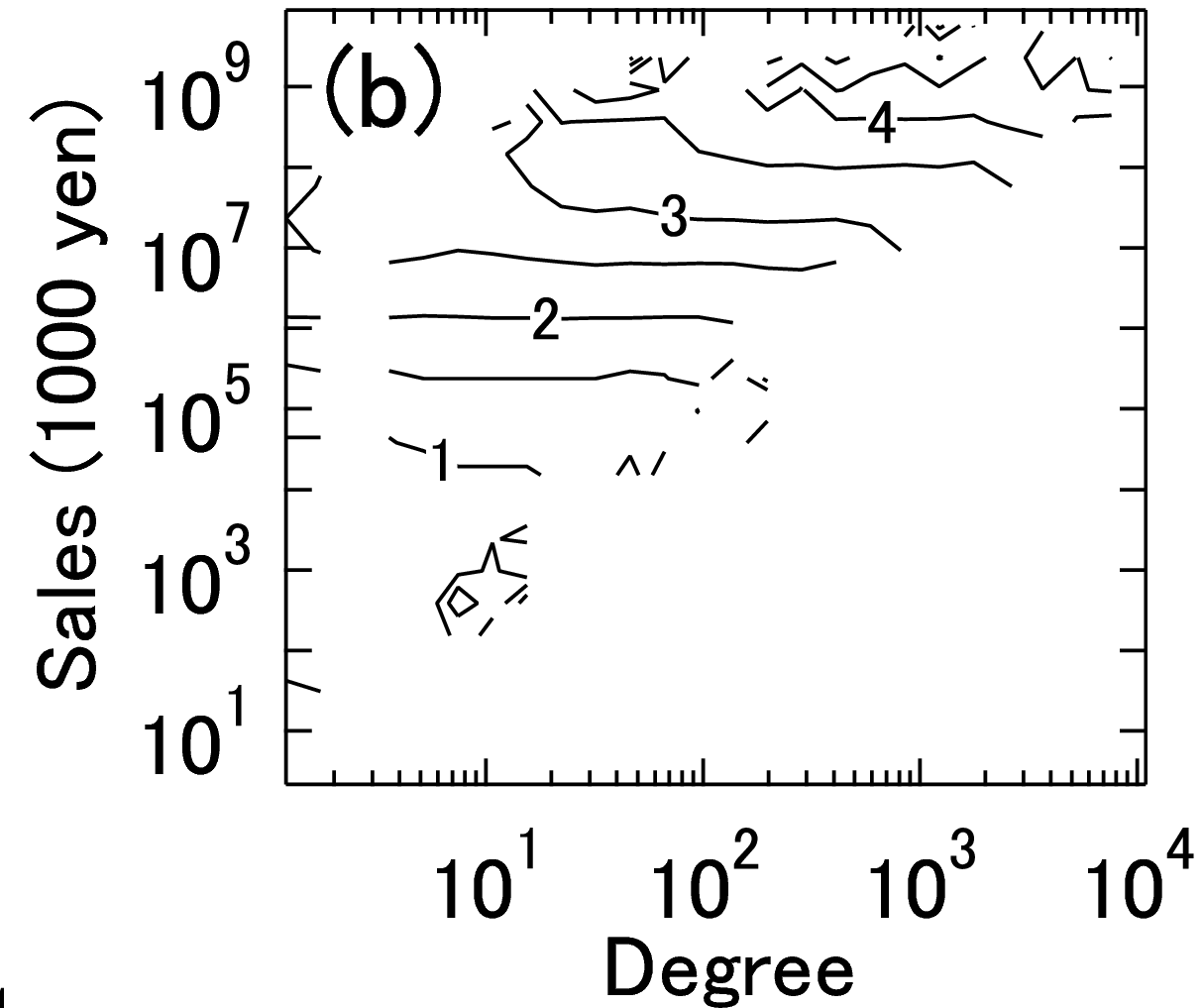}
\end{minipage}
\begin{minipage}{1\hsize}
\includegraphics[width=6.5cm]{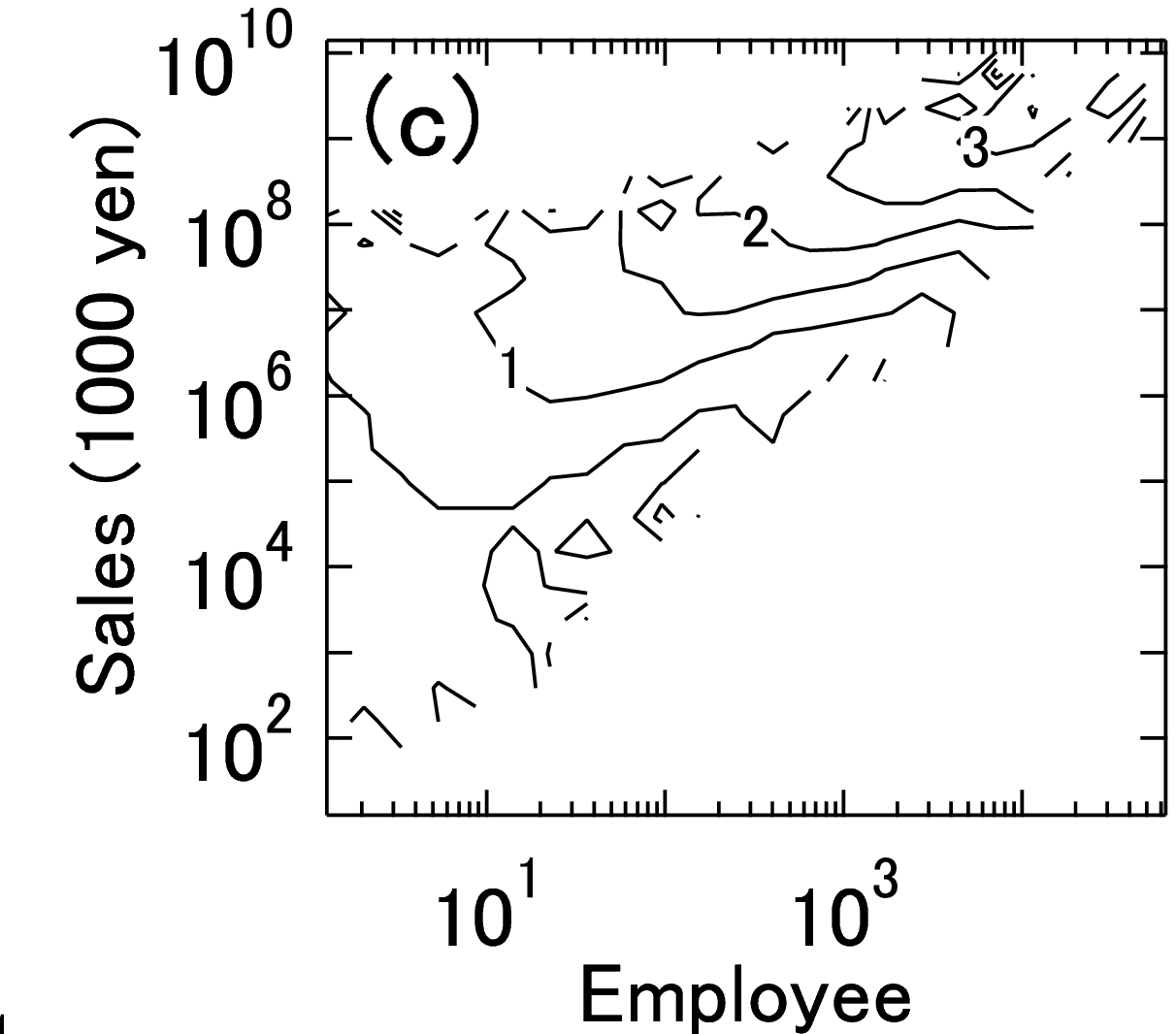}
\end{minipage}
\caption{
(a)Conditional median of sales $s$ given degree $k$ and number of employees $l$, $<s|k,l>_{0.5}$. The contour lines provide the sales s (1000 yen) in common logarithm (e.g., ``5" in the figure means $10^8$ yen). The red dashed line is $l \propto k^{0.44}$. 
(b)Conditional median of number of employees $l$ given values of degree $k$ and sales $s$, $<l|k,s>_{0.5}$. The contour lines provide $l$ in common logarithm. For example, ``3" in the figure means $l=10^{3}$. 
(c) Conditional median of degree $k$ given number of employees $l$ and sales $s$, $<k|l,s >_{0.5}$. The contour lines give $k$ in common logarithm.
}
\label{real_2d}
\end{figure}

\begin{figure*}
\begin{minipage}{0.45\hsize}
\includegraphics[width=8cm]{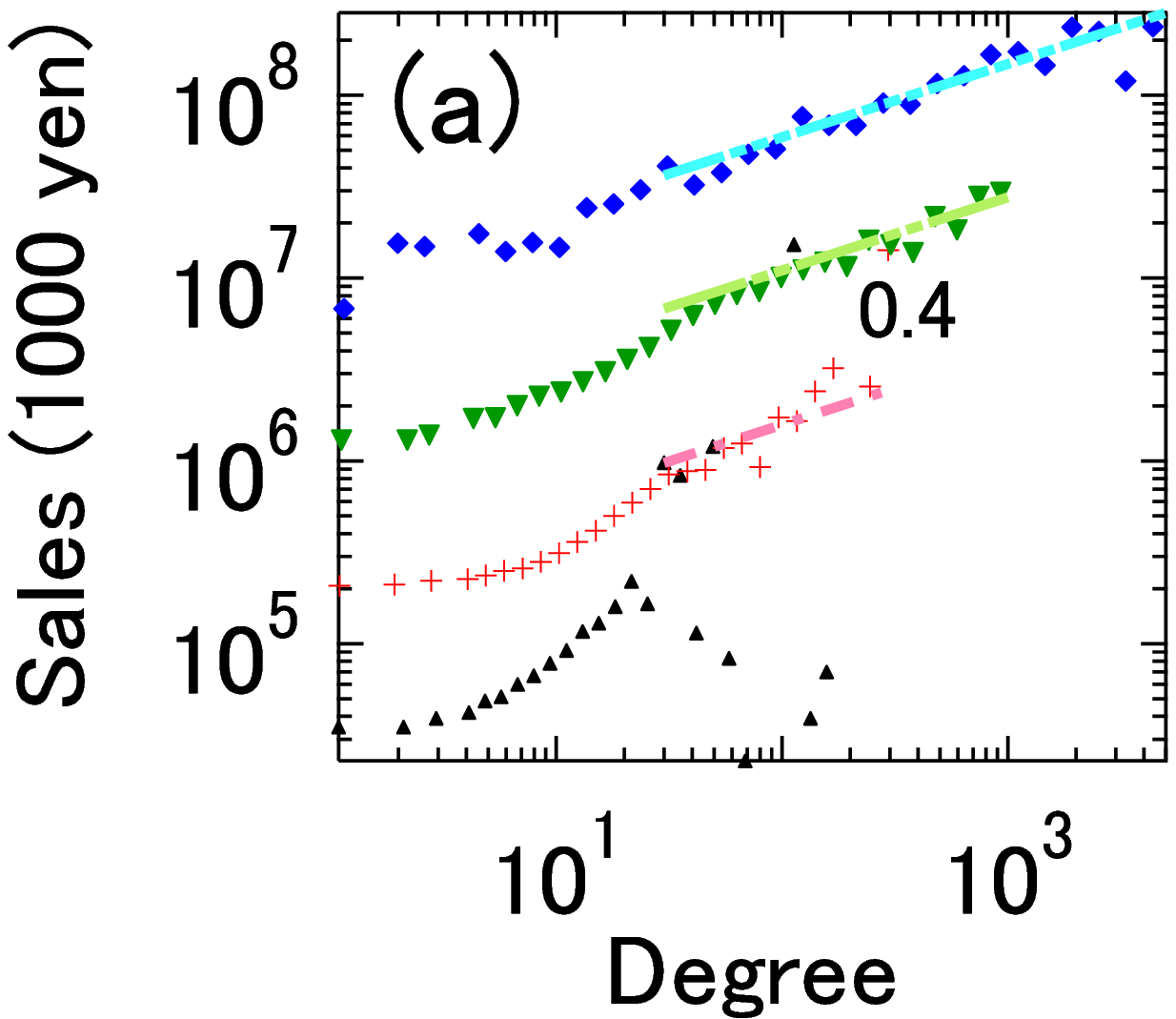}
\end{minipage}
\begin{minipage}{0.45\hsize}
\includegraphics[width=8cm]{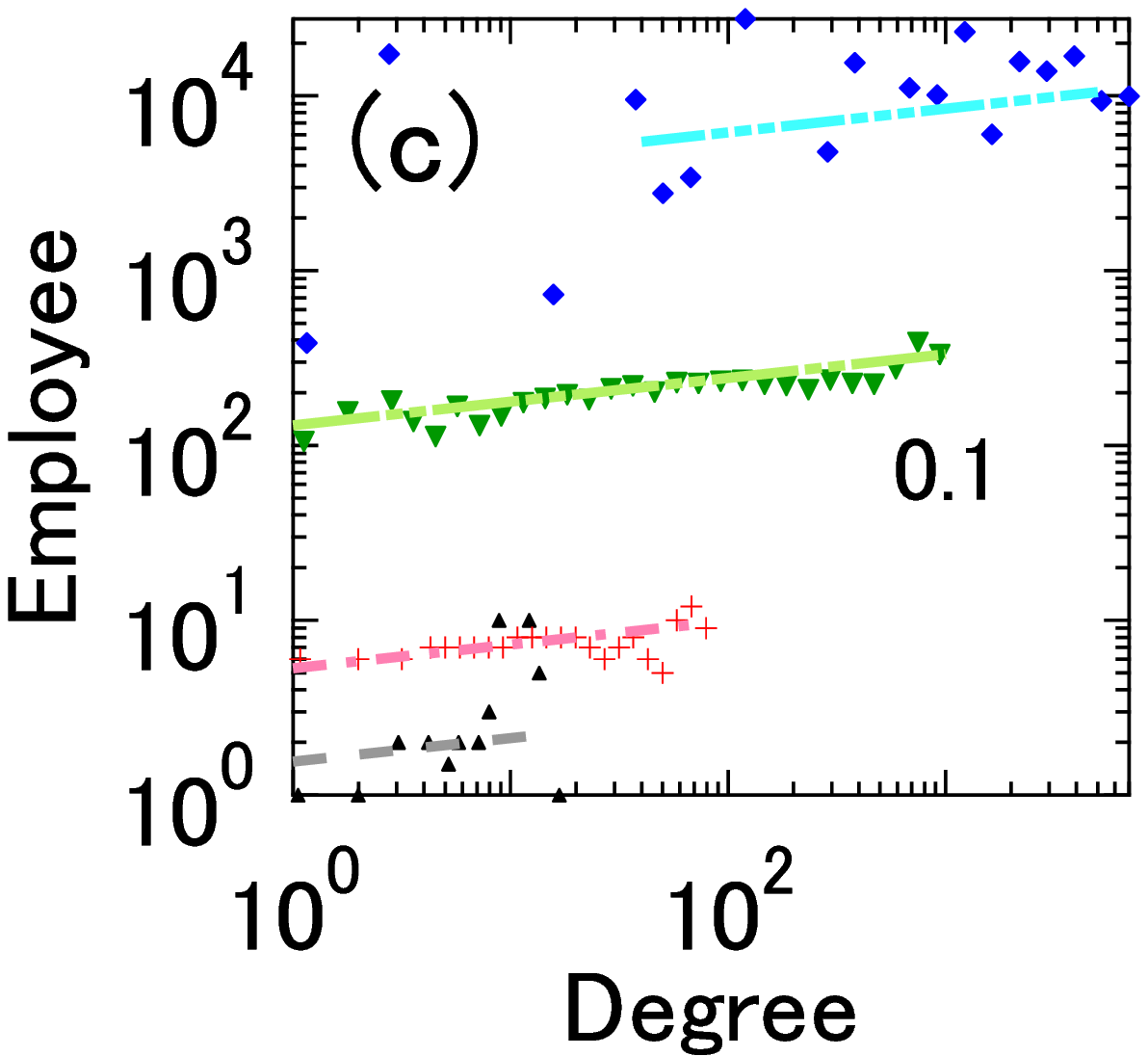}
\end{minipage}
\begin{minipage}{0.45\hsize}
\includegraphics[width=8cm]{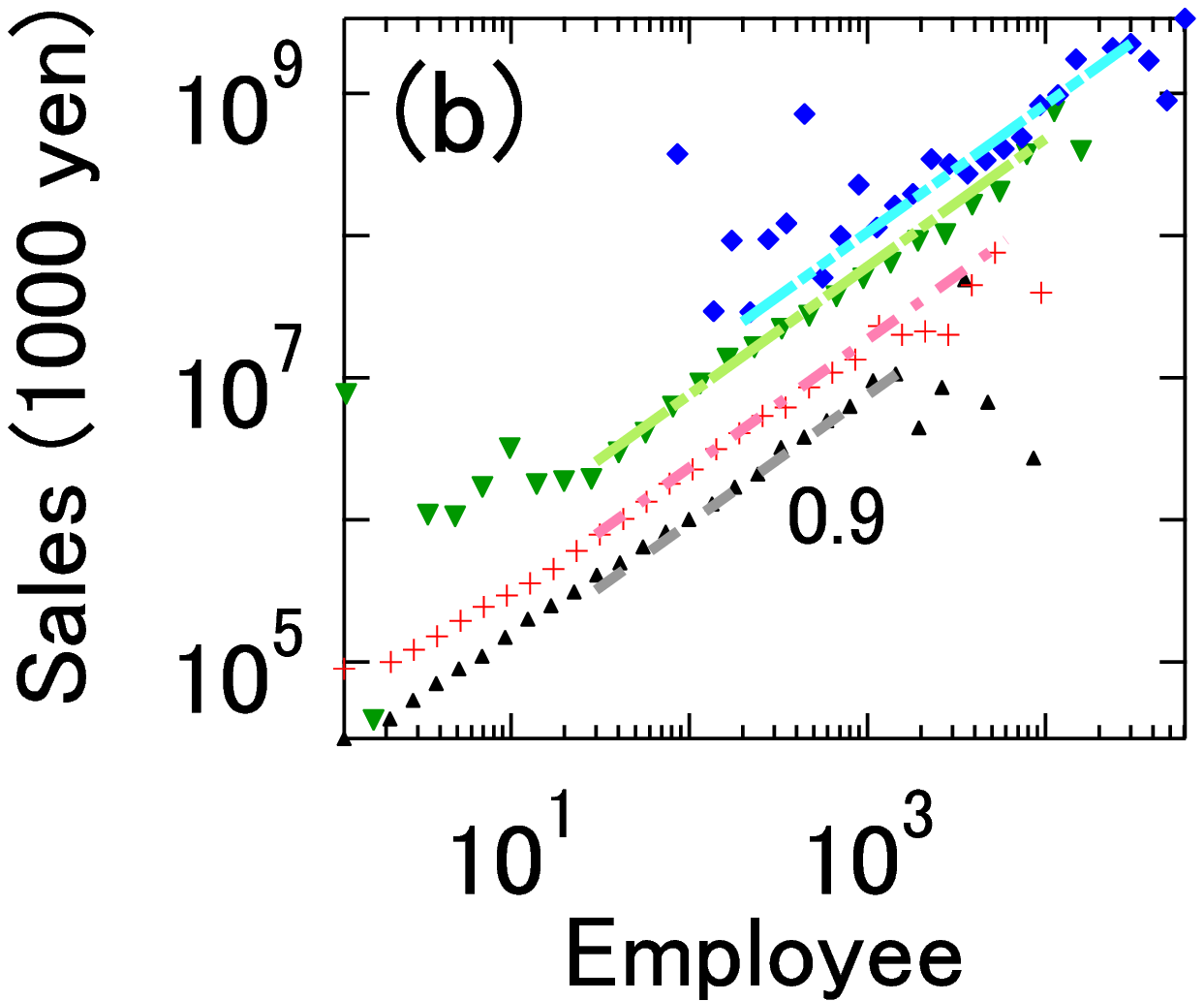}
\end{minipage}
\begin{minipage}{0.45\hsize}
\includegraphics[width=8cm]{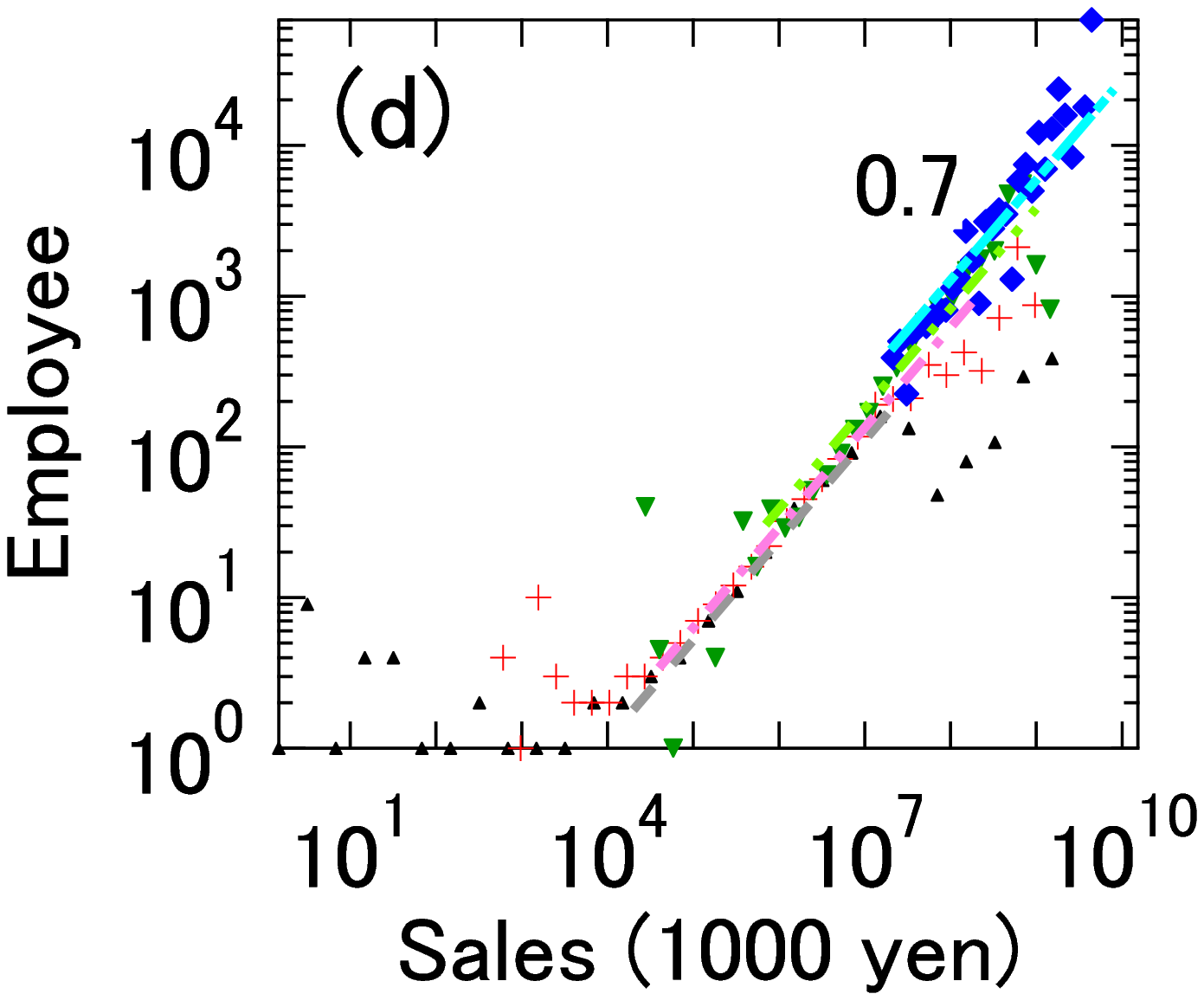}
\end{minipage}
\caption{
(a)Dependence of $<s|k,l>_{0.5}$ on sales $k$ regarding $l$ as fixing parameter for $1 \leq l < 2$ (black triangles), $10 \leq l <20$ (red pluses signs), $100 \leq l < 200$ (green nablas) and $1000 \leq l < 2000$ (blue diamonds). All supporting lines are proportional to $ \propto k^{0.4}$. \\
(b) Dependence of $<s|k,l>_{0.5}$ on sales $l$ regarding $k$ as fixing parameter for $1 \leq k < 2$ (black triangles), $10 \leq k <20$ (red pluses signs), $100 \leq k < 200$ (green nablas) and $1000 \leq k < 2000$ (blue diamonds).  All supporting lines are proportional to $ \propto l^{0.9}$.\\
(c)Dependence of $<l|k,s>_{0.5}$ on the employee $k$ regarding $s$ as fixing parameter for $1 \cdot 10^3 \leq s < 2 \cdot 10^3 $(1000 yen) (black triangles), $1\cdot 10^5 \leq s <2 \cdot 10^5$(1000 yen) (red pluses signs), $1\cdot 10^7 \leq s < 2 \cdot 10^7$(1000 yen) (green nablas) and $1\cdot 10^9  \leq s < 2 \cdot 10^9$ (1000 yen) (blue diamonds).  All supporting lines are proportional to $ \propto k^{0.1}$. \\
(d)Dependence of $<l|k,s>_{0.5}$ on the employee $s$ regarding $k$ as fixing parameter for $1 \leq k < 2$ (black triangles), $10 \leq k <20$ (red pluses signs), $100 \leq k < 200$ (green nablas) and $1000 \leq k < 2000$ (blue diamonds). All supporting lines are proportional to $ \propto s^{0.7}$.}
\label{real_1d}
\end{figure*}

\subsection{Correlations among three variables}
In this subsection, we investigate the dependence of a given variable on the others. 
Here we plot only the median values of $X$ to characterize of the conditional probability density $P(X|Y,Z)$ because the number of observable 
samples is not sufficiently large by conditioning two variables $Y$ and $Z$. 
Although the mean value is another candidate for characterizing the probability density, the median value is much more robust than the mean value for outliers, thus we use the median value because the data we are analyzing include outliers. \par 

Fig. \ref{real_2d} shows contour plots of $<s|k,l>_{0.5}$, $<l|k,s>_{0.5}$, and $<k|l,s>_{0.5}$, 
where $<X|Y,Z>_{0.5}$ is the conditional median of $X$ for the given values of $Y$ and $Z$. 
The contour lines of $<s|k,l>_{0.5}$ are characterized by oblique lines with a slope of $0.44$, as shown in Fig. \ref{real_2d}(a), 
whereas the contour lines of $<l|k,s>_{0.5}$ are approximated by almost-horizontal lines, as shown in Fig \ref{real_2d}(b). 
Similarly, the contour plots of $<k|l,s>_{0.5}$ are shown in Fig. \ref{real_2d}(c), which is clearly different from the former two cases. \par  
To better understand of these correlations, we investigate the dependence of the conditional median for each variable. 
Fig. \ref{real_1d}(a) shows how $<s|k,l>_{0.5}$ depends on degree $k$ when we regard $l$ as a fixed parameter. 
From this figure, we find that the value of $<s|k,l>_{0.5}$ is characterized by a power law with base $k$ and exponent $0.4$.  
From Fig. \ref{real_1d}(b) we see that $<s|k,l>_{0.5}$  depends on the degree $l$ when we regard $k$ as a fixed parameter.  
We find that $<s|k,l>_{0.5}$ is proportional to $l^{0.9}$. 
Combining these two results, we have the following scaling law:
\begin{equation}
<s|k,l>_{0.5} \propto k^{{\gamma_{s|k,l}}^{(k)}} \cdot l^{{\gamma_{s|k,l}}^{(l)}}. \label{pro1}
\end{equation}
where ${\gamma_{s|k,l}}^{(k)}=0.4$ and ${\gamma_{s|k,l}}^{(l)}=0.9$. \par
 Fig. \ref{real_1d}(c) shows how $<l|k,s>_{0.5}$ depends on the degree $k$ when we regard $s$ as a fixing parameter. 
From this figure, we can see that $<l|k,s>_{0.5}$ is proportional to $k^{0.1}$. Similarly, from Fig. \ref{real_1d}(d), we find that $<l|k,s>_{0.5}$ is proportional to $s^{0.7}$. Thus, we have the following scaling law:
\begin{equation}
<l|k,s>_{0.5} \propto k^{{\gamma_{l|k,s}}^{(k)}} \cdot s^{{\gamma_{l|k,s}}^{(s)}} \label{pro2},
\end{equation}
where ${\gamma_{l|k,s}}^{(k)}=0.1$ and  ${\gamma_{l|k,s}}^{(s)}=0.7$.
Note that the non-trivial scaling relations,  Eqs. (\ref{pro1}) and (\ref{pro2}), can be derived by carefully analyzing the conditional statistics of three variables. \par 

\section{The models}

\begin{figure*}
\begin{minipage}{0.45\hsize}
\includegraphics[width=8cm]{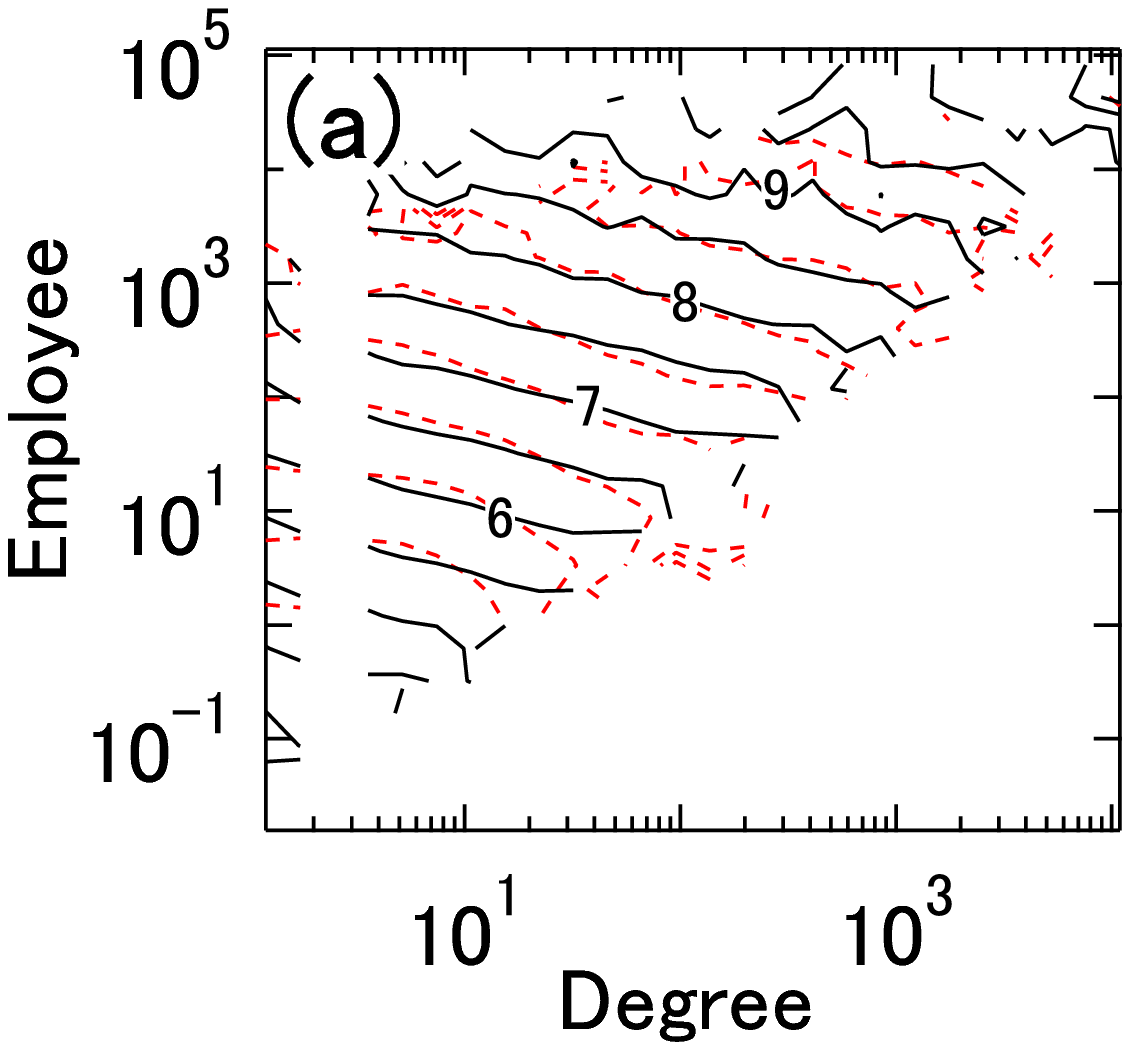} 
\end{minipage}
\begin{minipage}{0.45\hsize}
\includegraphics[width=8cm]{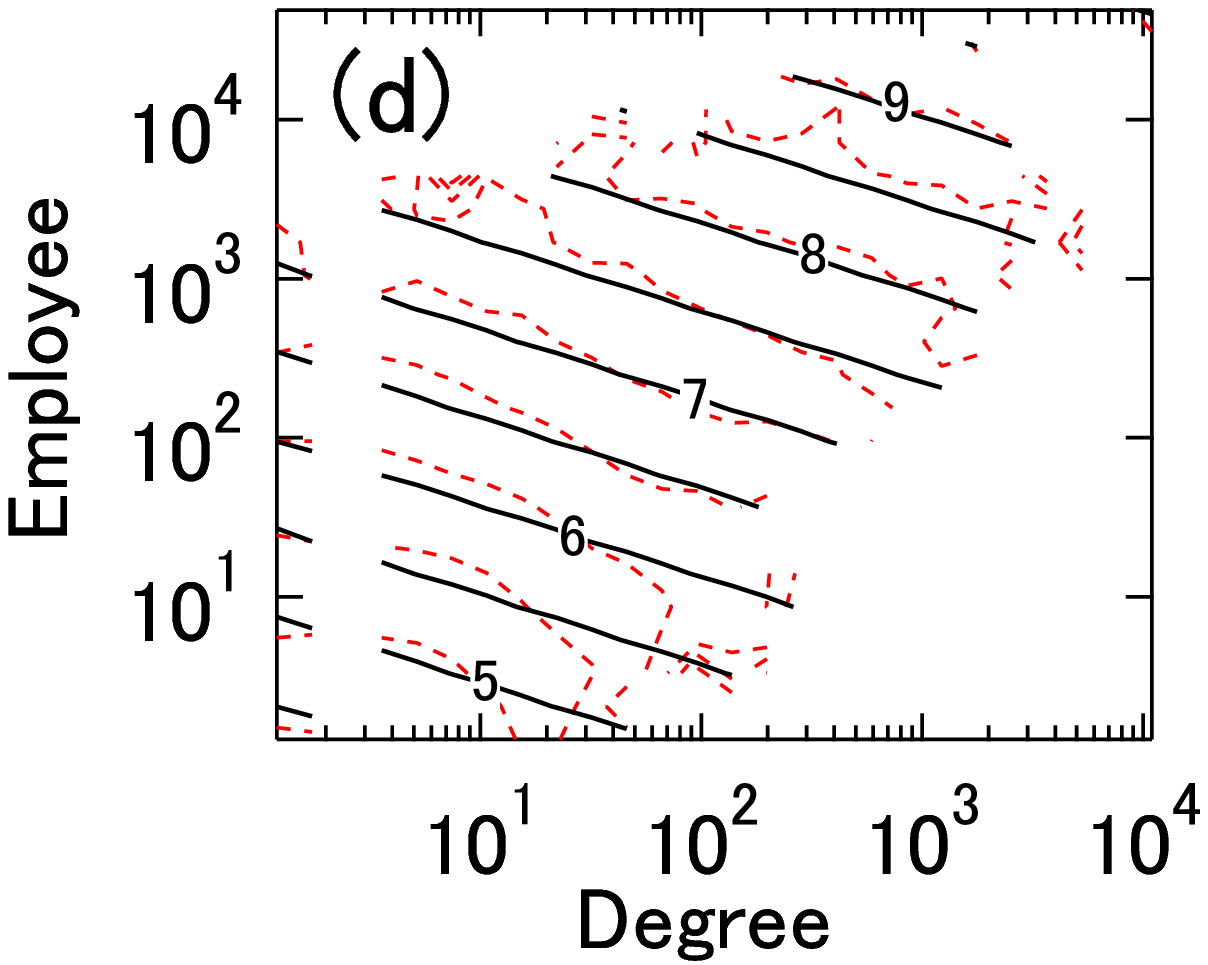}
\end{minipage}
\begin{minipage}{0.45\hsize}
\includegraphics[width=8cm]{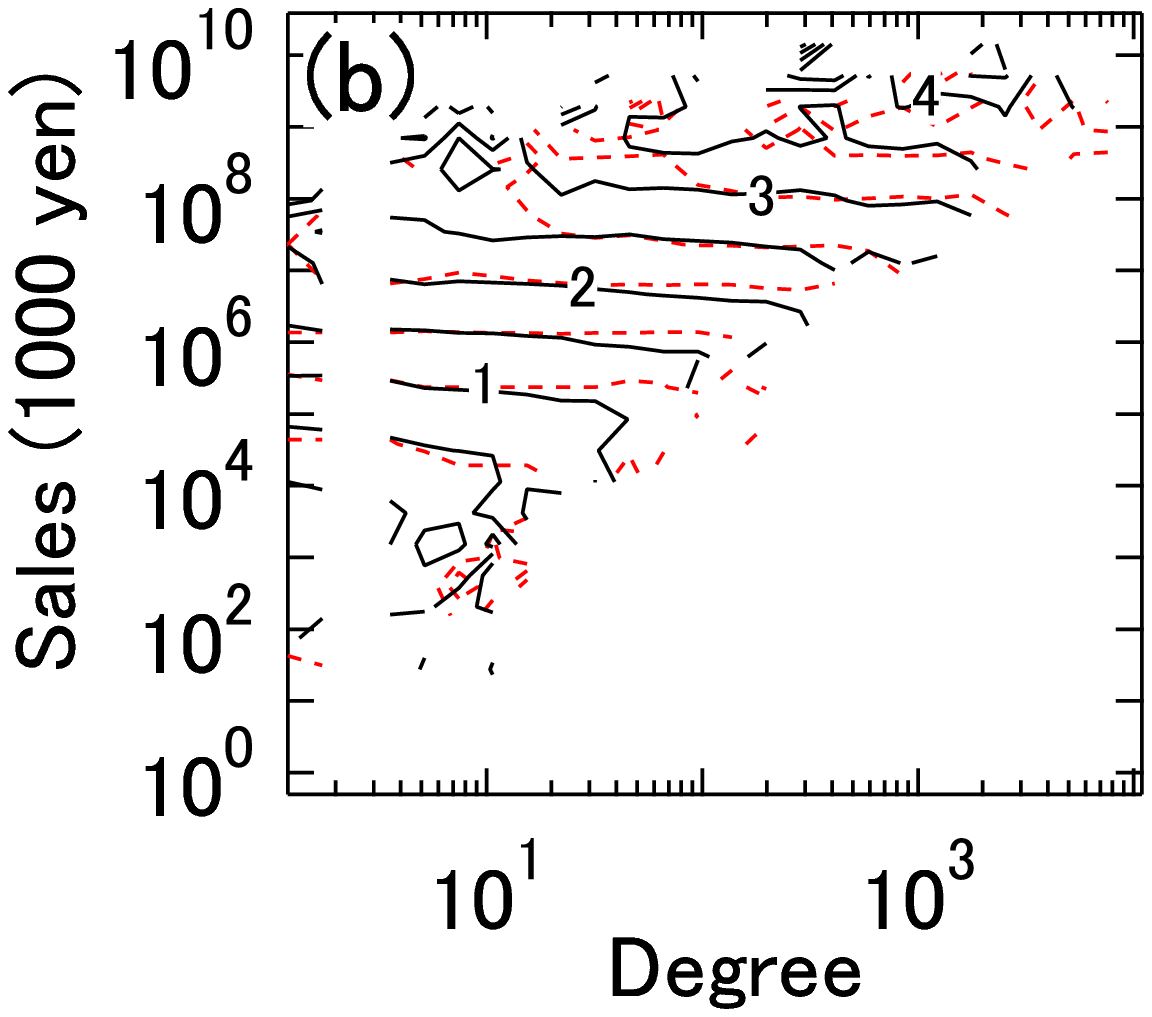}
\end{minipage}
\begin{minipage}{0.45\hsize}
\includegraphics[width=8cm]{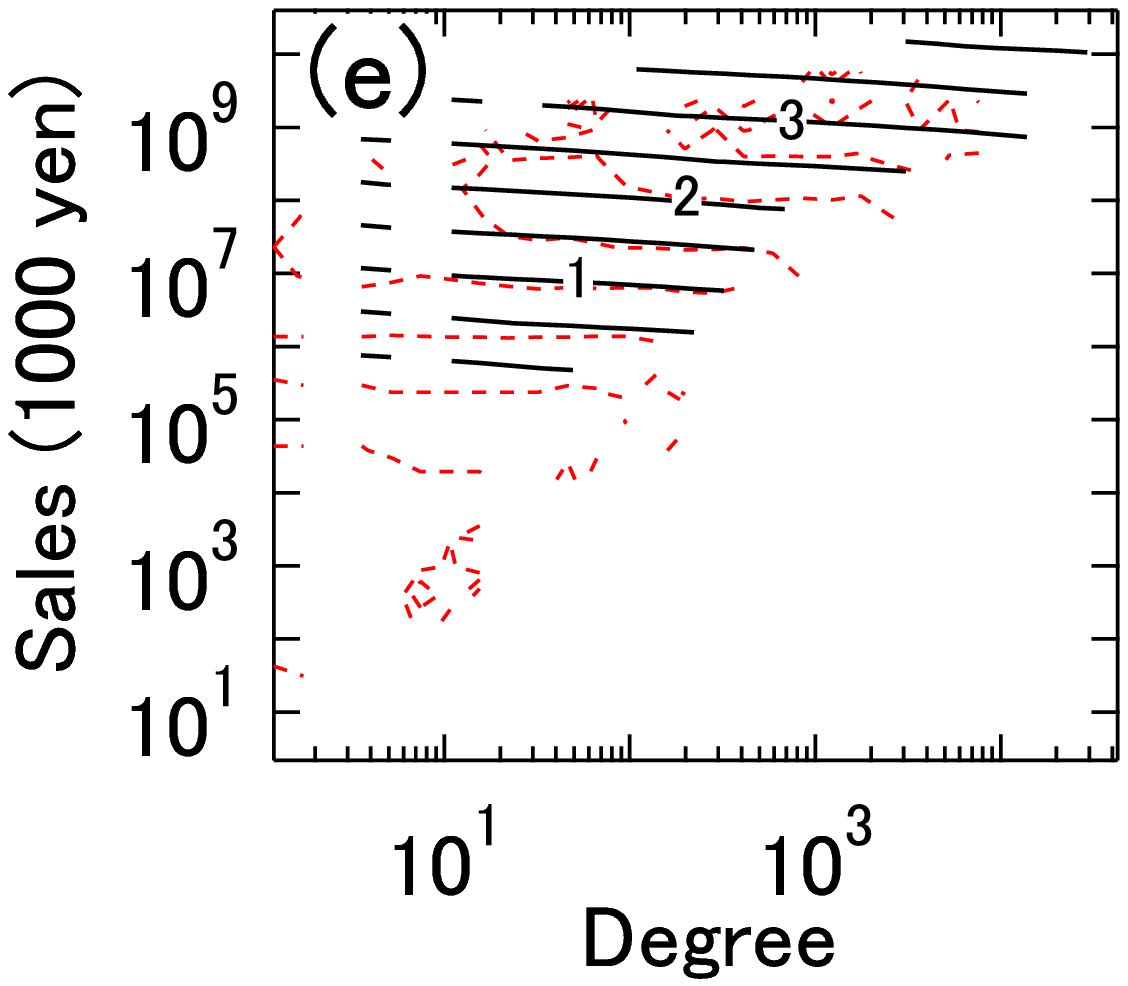}
\end{minipage}
\begin{minipage}{0.45\hsize}
\includegraphics[width=8cm]{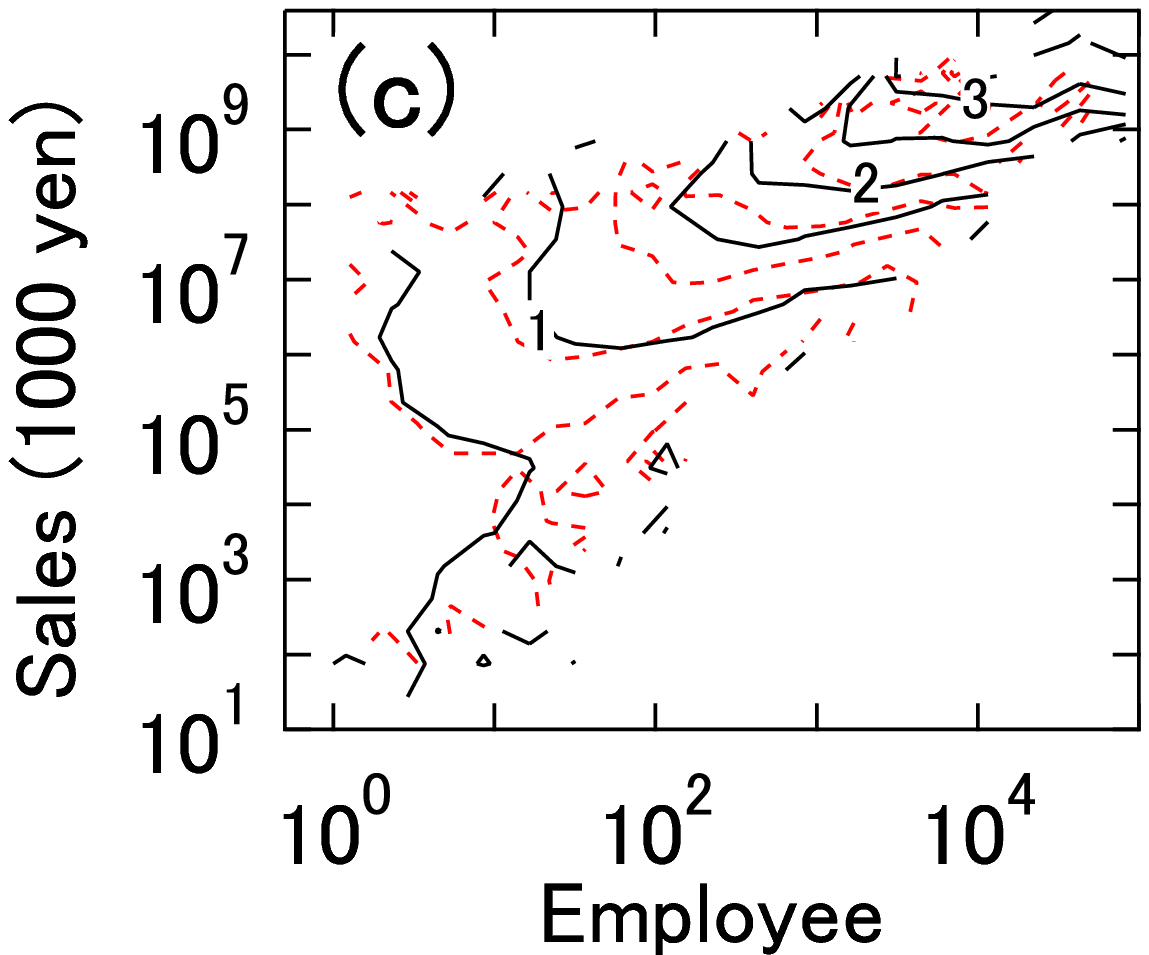}
\end{minipage}
\begin{minipage}{0.45\hsize}
\includegraphics[width=8cm]{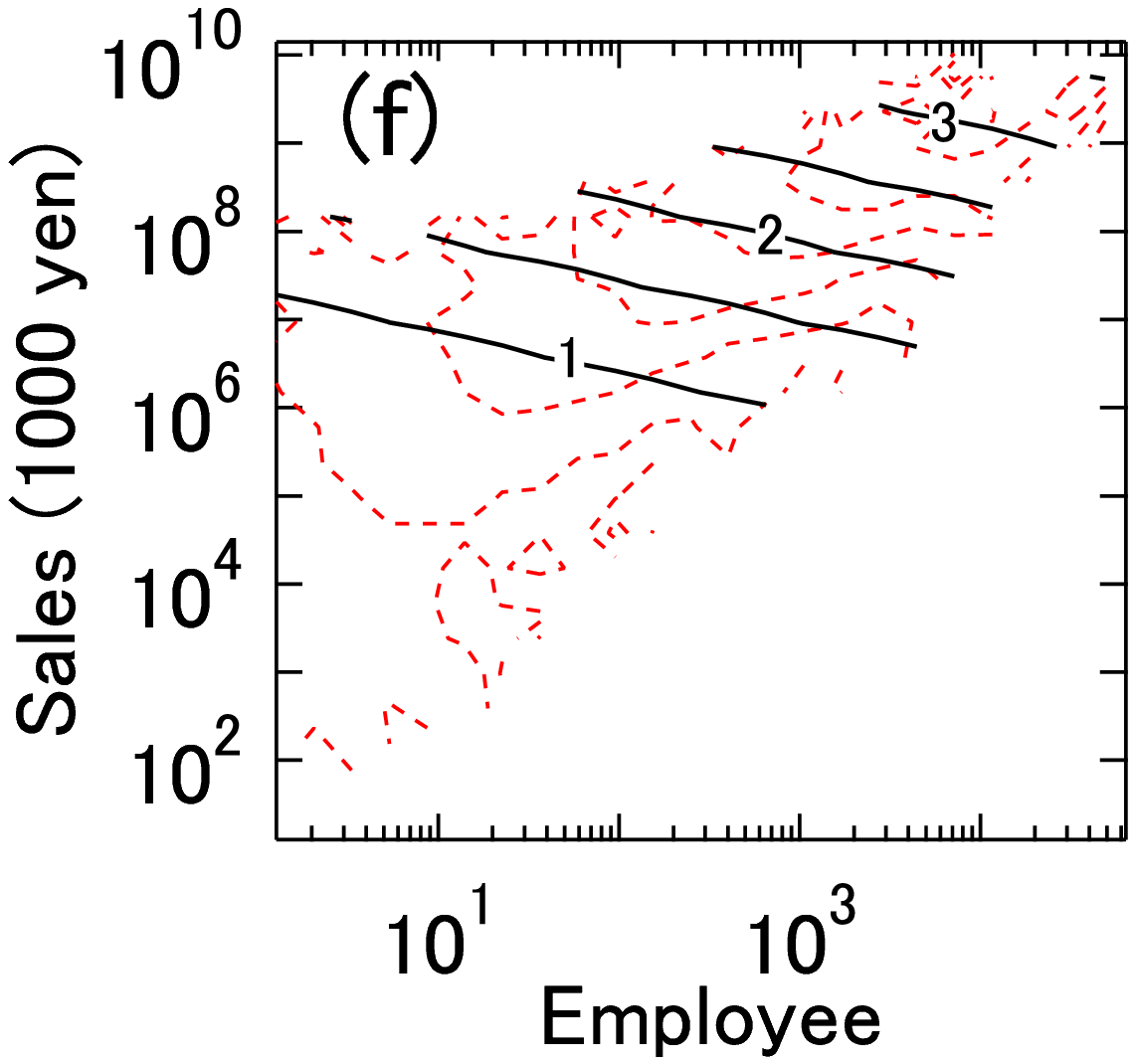}
\end{minipage}
\caption{(a) Conditional median of sales $s$ for given degree $k$ and number of employees $l$, $<s|k,l>_{0.5}$, for shuffled model. 
The black lines are the contour lines for a given $s$ (the numbers on the lines denote the digits of annual sales in multiples of 1000 yen; for example, ``5" in the figure means $10^{8}$ yen). The red dashed line provides the corresponding contours for actual data (we plot the other cases in a similar manner). (b) Conditional median of the number of employees $l$ for given degree $k$ and sales $s$, $<l|k,s>_{0.5}$, for shuffled model. 
The contour lines of $l$ are plotted in a similarly manner as for panel (a)
 (the numbers on the lines show the digits; for example, ``3" in the figure means $l = 10^3$). 
 (c)Conditional median of degree $k$ given by number of employees $l$ and sales $s$, $< k|l,s >_{0.5}$, for shuffled model. 
 (d)-(f) Panels corresponding to panels (a)-(c), respectively, for lognormal distribution model. }
\label{shuff_2d}
\end{figure*}

\begin{figure}
\begin{minipage}{1\hsize}
\includegraphics[width=7.6cm]{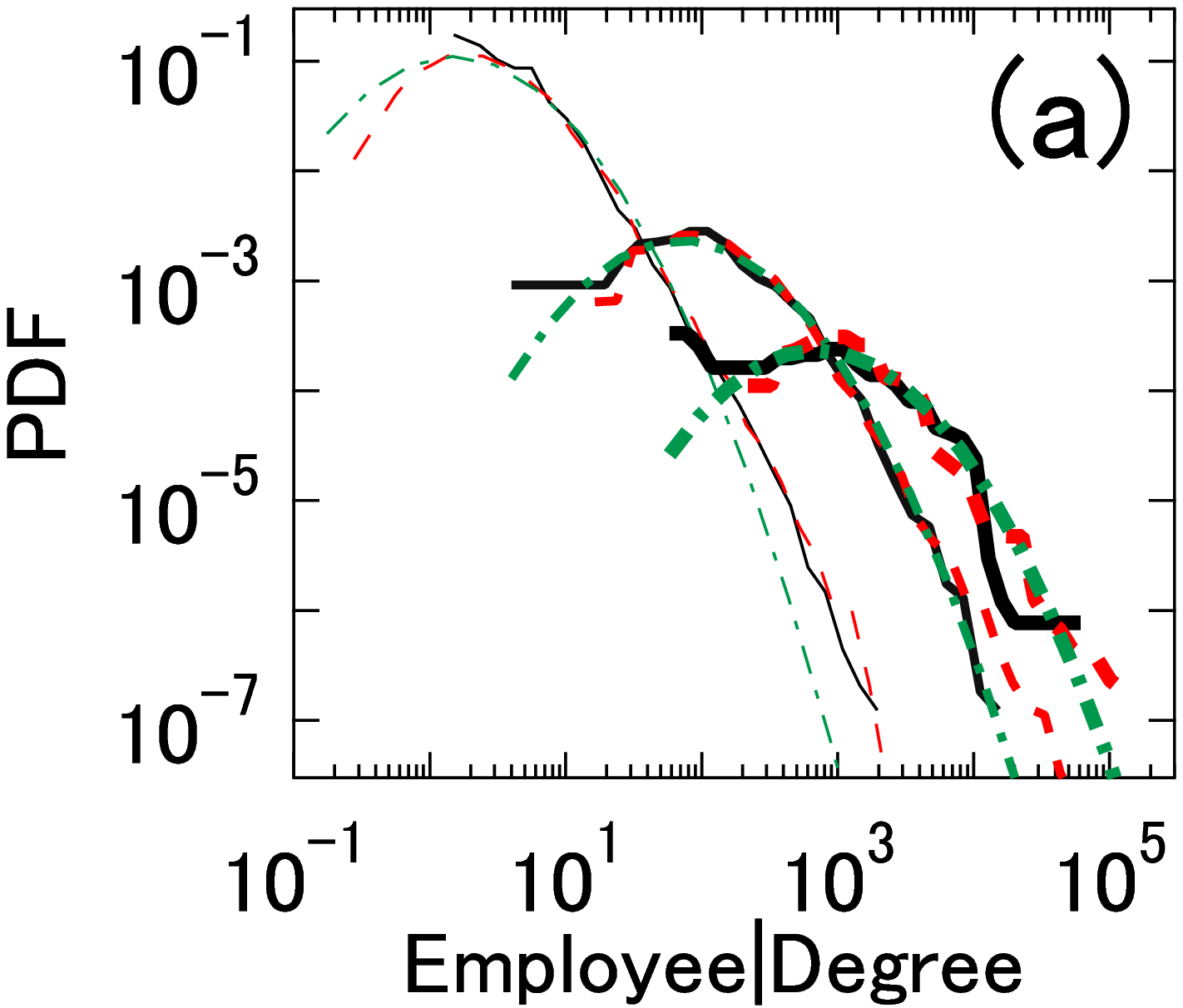}
\end{minipage}
\begin{minipage}{1\hsize}
\includegraphics[width=7.6cm]{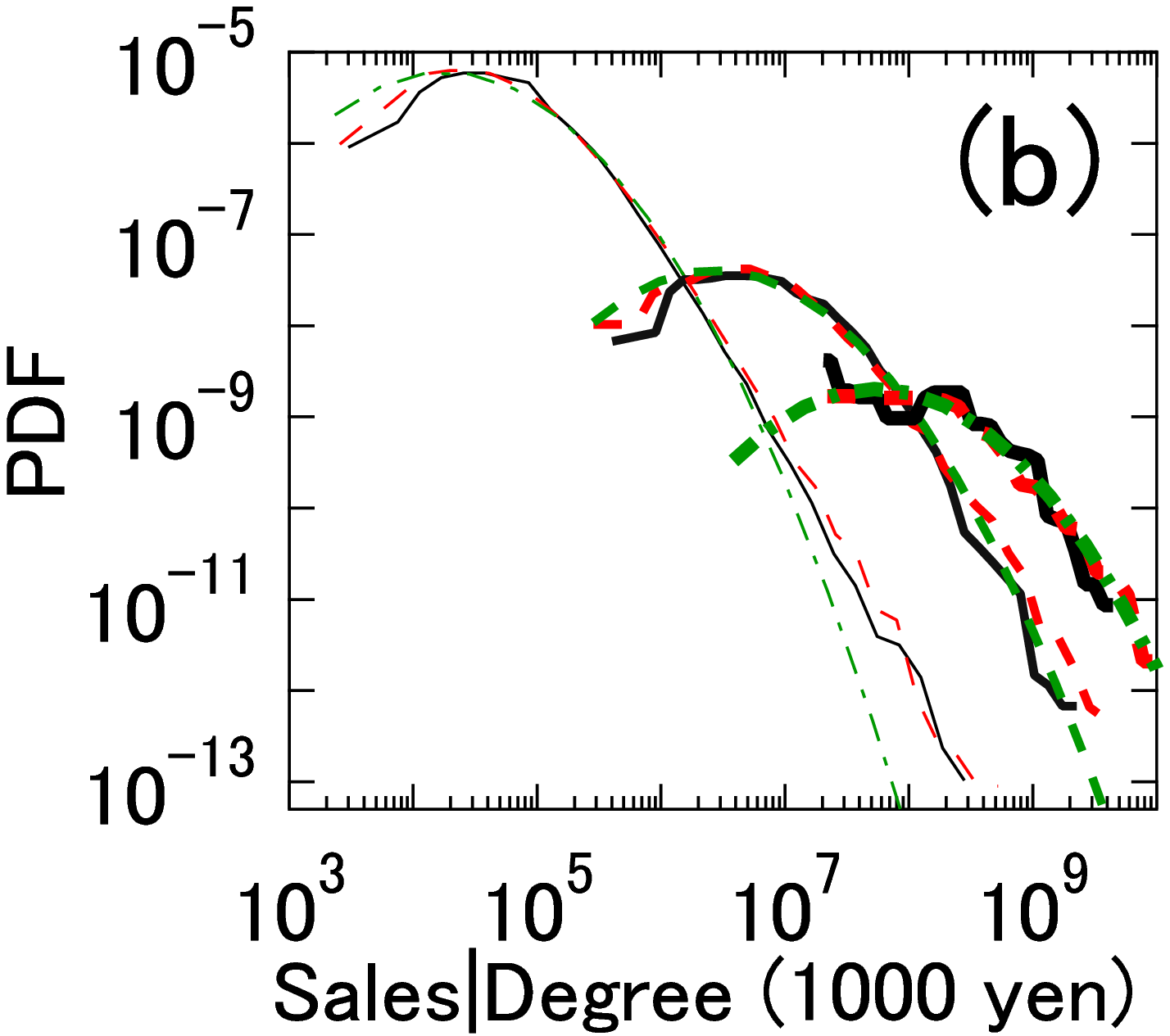}
\end{minipage}
\begin{minipage}{1\hsize}
\includegraphics[width=7.6cm]{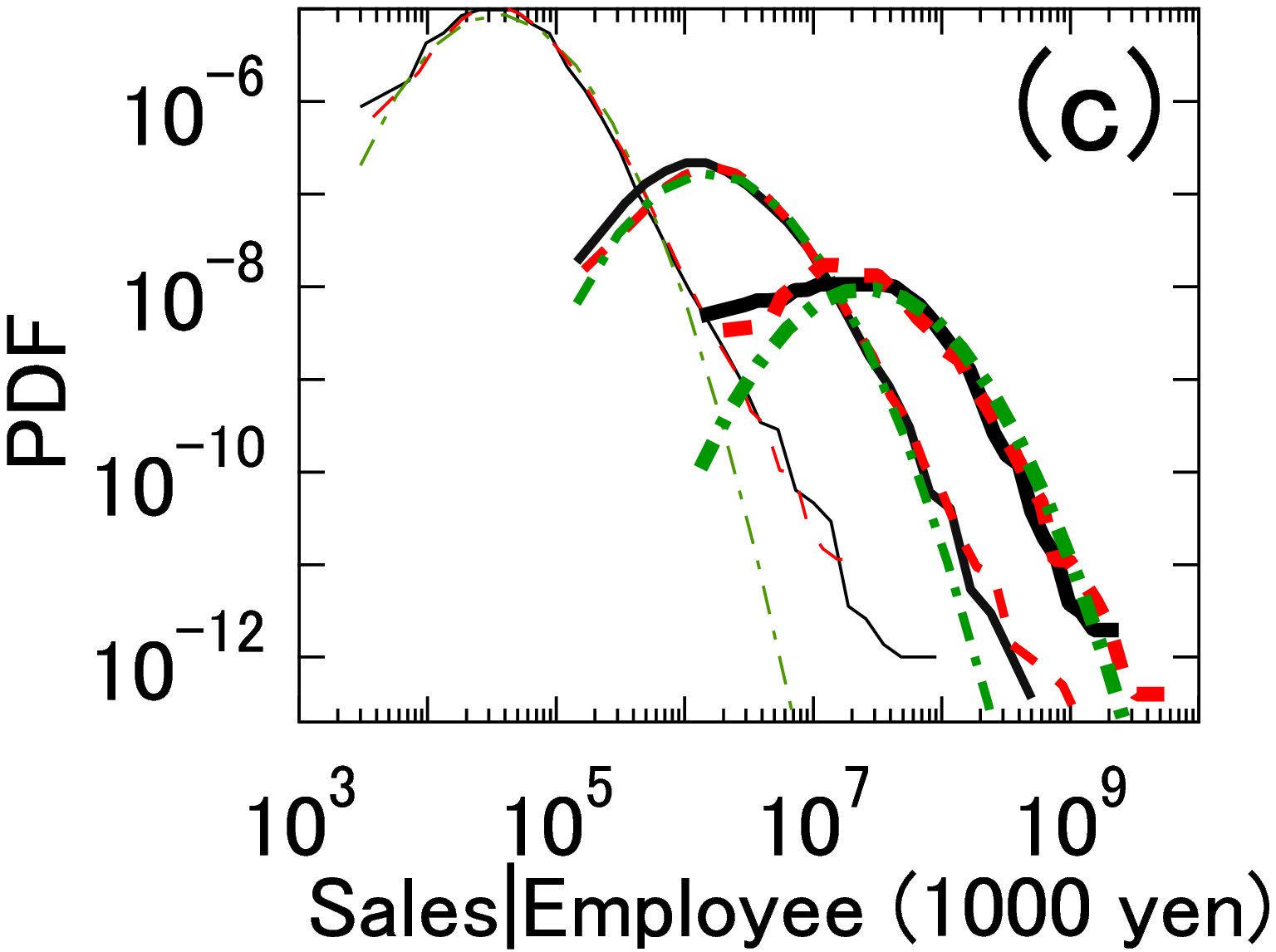}
\end{minipage}
\caption{Comparison of conditional PDFs, $P(l|k)$, $P(s|k)$, and $P(s|l)$ for actual data (black solid lines), shuffle model (red dashed lines) and lognormal distribution model (green dash-dotted lines). 
(a)Conditional PDF of employee $l$ for given degree $k$, $P(l|k)$ for $k=3$ (thin lines), $k=14 \cdot 10$ (medium lines) and $k=14 \cdot 10^2$ (thick lines). For actual data and the shuffle model, we apply the following conditions  to include a sufficient number of samples: $k=14 \cdot 10 \approx \sqrt{100 \cdot 200}$ for the interval $100 \leq k < 200$ and $k=14 \cdot 10^2 \approx \sqrt{1000 \cdot 2000}$ for the interval $1000 \leq $k$ < 2000$. For the lognormal distribution model, the PDF is given by Eq. (\ref{l_k_p_a}). 
(b)Corresponding plots for conditional PDFs of sales $s$ for given $k$, $P(s|k)$. For the lognormal distribution model, the PDF is given by Eq. (\ref{s_k_p_a}).
 (c)Conditional PDFs of sales $s$ for given $l$, $P(s|l)$, for $l=3$ (thin lines), $l=14 \cdot 10$ (medium lines) and $l=14 \cdot 10^2$ (thick lines). For actual data and the shuffle model, we apply the following conditions: $l=14 \cdot 10 \approx \sqrt{100 \cdot 200}$ for the interval $100 \leq l < 200$ and $l=14 \cdot 10^2 \approx \sqrt{1000 \cdot 2000}$ for the interval $1000 \leq $l$ < 2000$. For the lognormal distribution model, the PDF is given by Eq. (\ref{s_l_p_a}). }
\label{syuuhen2}
\end{figure}

\begin{figure}
\begin{minipage}{1\hsize}
\includegraphics[width=8cm]{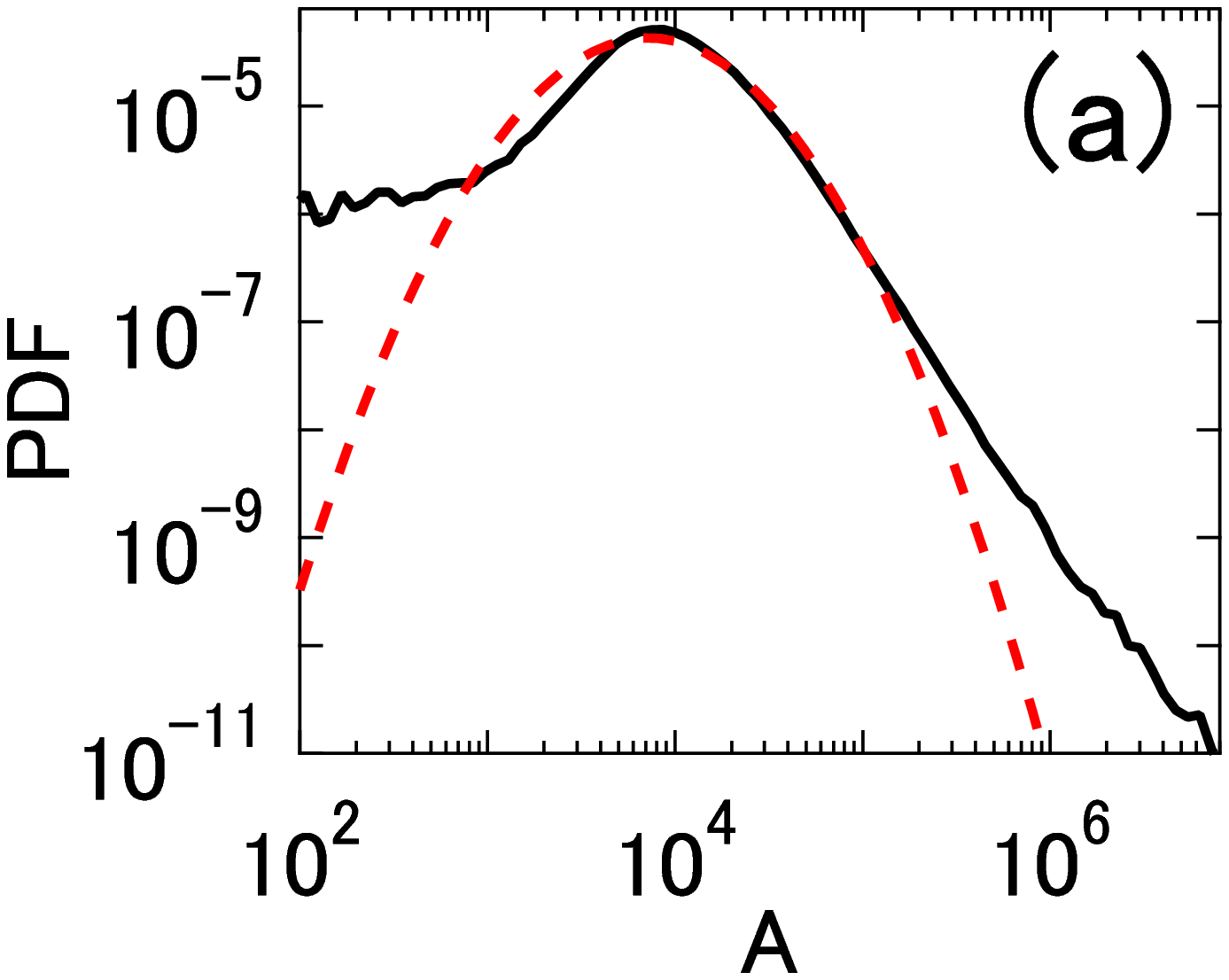} 
\end{minipage}
\begin{minipage}{1\hsize}
\includegraphics[width=8cm]{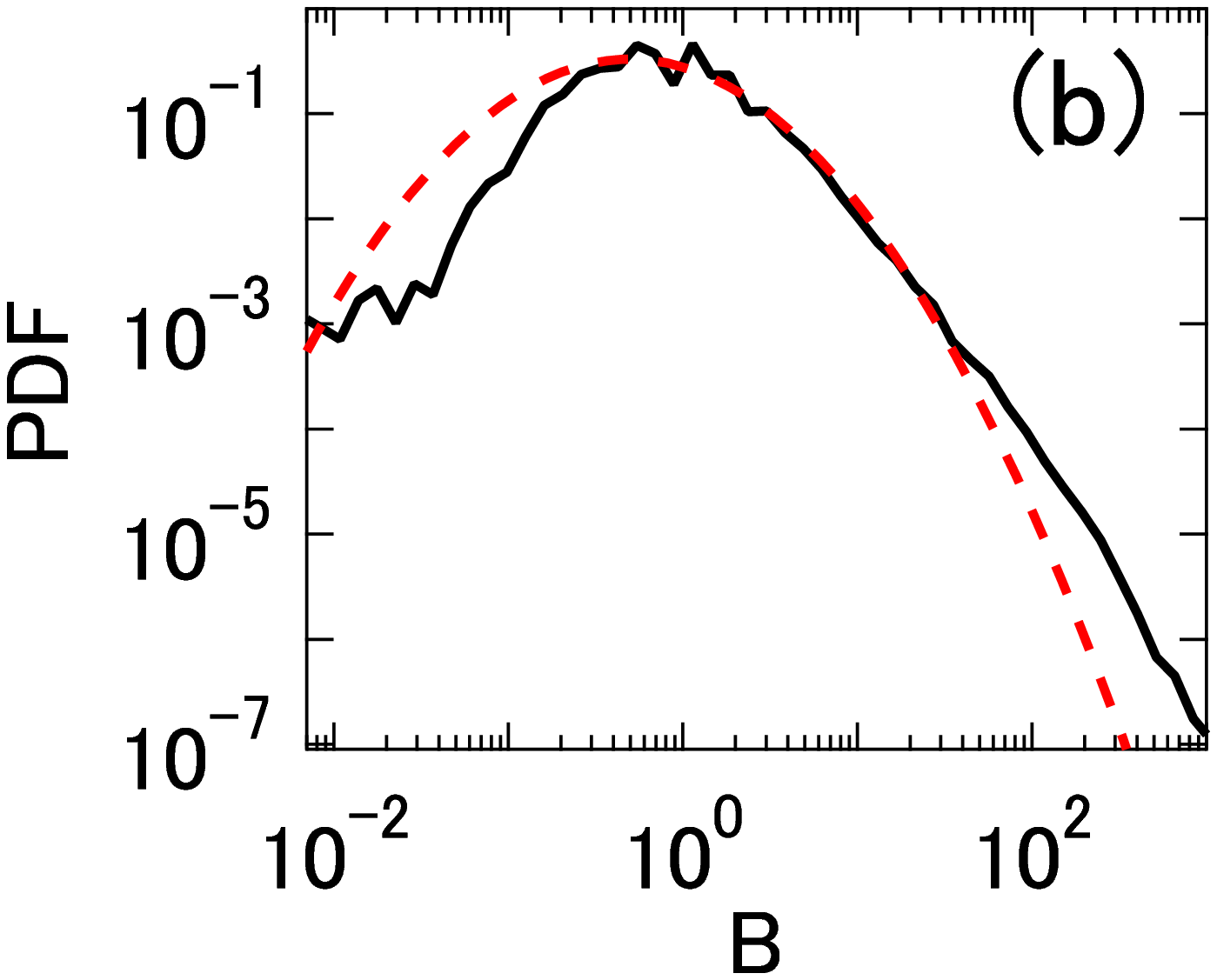} 
\end{minipage}
\caption{(a)PDF of coefficient. 
$A$ given by Eq. (\ref{model3}) for actual data (black solid line: by definition the shuffle model gives the same PDF), and for the lognormal distribution model (red dashed line). 
(b)PDF of $B$ given by Eq. (\ref{model2}) for actual data (black solid line: by definition the shuffle model gives the same PDF) and for the lognormal distribution model (red dashed line). For both cases,  the central parts of the real distributions are approximated by the lognormal distribution model. }
\label{theory}
\end{figure}
To clarify any mutual relation between the above-mentioned empirical scalings, 
we now introduce some simple models.
First, note that the empirical relations, Eqs. (\ref{pro1}) and (\ref{pro2}), seem inconsistent if we neglect fluctuations. 
For example, if we assume Eq. (\ref{pro1}), $s \propto k^{0.4} \cdot l^{0.9}$, 
then we have $l \propto k^{-0.44} s^{1.1}$. 
However, this result disagrees with the empirical scaling given by Eq. (\ref{pro2}).  
Therefore, to reproduce the empirical observations, 
we must take into account the effects of fluctuations, which modify the scaling relations.
Here, we introduce a simple model involving $k$, $l$, and $s$ that assumes that these variables are
 derived from three independent random variables $K$, $B$, and $A$ as follows:  
\begin{equation}
k=K \label{model1}
\end{equation}
\begin{equation}
l=B \cdot k \label{model2}
\end{equation}
\begin{equation}
s=A \cdot l^{\alpha} \cdot k^{\beta} \label{model3},
\end{equation}
where Eqs. (\ref{model2}) and (\ref{model3}) refer to Eqs. (\ref{l_k_kika}) and (\ref{pro1}) respectively. 
Here, $\alpha=0.9$ and $\beta=0.4$. \par  
In this model, we determine $k$, $l$ and $s$ in the following order:
\begin{enumerate}
\item We determine the degree $k$ by sampling random variables, which we specify in the following discussion of $K$.
\item We determine the employee $l$ from Eq. (\ref{model2}) and by using the degree $k$ determined in the previous step, where the value of $B$ is determined by a random variable.
\item We determine $s$ from Eq. (\ref{model3}) and using $k$ and $l$ determined in the previous steps, where the value of $A$ is determined by a random variable. 
\end{enumerate}

\subsection{Shuffled model}
First, let us introduce the model in which we choose random variables from the real values by using a bootstrapping method.
 We randomly resample $K$ from shuffled actual degrees $k_i \quad (i=1,2,\cdots,N)$, and $B$ and $A$ are similarly chosen randomly from shuffled actual data, $l_i/k_i \quad (i=1,2,\cdots,N)$,  $s_i/(k_i^{\beta} \cdot l_i^{\alpha}) \quad (i=1,2,\cdots,N)$ respectively. 
Figs. \ref{shuff_2d}(a), (b), and (c), show the comparisons between simulation results and actual results with respect to $<s|l,k>_{0.5}$, $<l|k,s>_{0.5}$, and  $<k|l,s>_{0.5}$. 
The results shown in there figures confirm that the model shown by black solid lines almost reproduces contours of the actual data, which are shown by red dashed lines.
In addition, we also confirm that the model reproduces the conditional probabilities between two variables, $P(l|k)$, $P(s|k)$, and $P(s|l)$, which are shown by the red dashed line in Fig. \ref{syuuhen2}, 
and the marginal distributions $P(k)$, $P(l)$, and $P(s)$ shown by the red dashed line in Fig. \ref{sed}. 
In all cases the distributions are nicely reproduced by this shuffled model. \par

The differences between this model and actual phenomena as follows: (i)The correlation between $B$ and $k$ is removed. (ii)The correlation between $A$ and $l^{\alpha}k^{\beta}$ is removed. (iii)With respect to Eqs. (\ref{model2}) and (\ref{model3}), there are non-power-law regions for small values for the case of actual observations. However, we approximate the single power laws for all regions of the model for simplification. (iv)Discrete quantities for actual data are approximated by continuous quantities.

In addition, in general, this model is not only the one that can reproduce empirical scaling relations, for example, we can change the order of the variables. 
By checking all combinations we find that this model with the given order of construction produces most accurate results upon comparing with the real data in our framework. 
This simple reconstruction model is based on the definition of conditional probability for three-body stochastic variables and uses the empirically derived scaling relations for the conditional probability densities. \par

\subsection{Lognormal distribution model}
\label{sec:theory}
\begin{table}
\begin{tabular}{cc ccc}
\hline \hline 
Scaling & Exponent&Theory &Value&Figure\\  \hline \hline 
$P(k) \propto k^{-\zeta_k-1}$ & $\zeta_k$ & $\lambda$ & 1.3 & \ref{sed}(a) \\ \hline
$P(l) \propto k^{-\zeta_l-1}$ & $\zeta_l$ & $\lambda$ & 1.3 & \ref{sed}(b) \\ \hline 
$P(s) \propto k^{-\zeta_s-1}$ & $\zeta_s$ & $\frac{\lambda}{\alpha+\beta}$& 1.0& \ref{sed}(c) \\ \hline
$<l|k>_{0.5} \propto k^{\gamma_{l|k}}$ & $\gamma_{l|k}$ & $1$ & 1 & \ref{degree_employee}(b) \\  \hline 
$<s|k>_{0.5} \propto k^{\gamma_{s|k}}$ & $\gamma_{s|k}$ & $\alpha+\beta$ & 1.3 & \ref{degree_sales}(b) \\ \hline  
$<s|l>_{0.5} \propto k^{\gamma_{s|l}}$ & $\gamma_{s|l}$ & $\alpha+\beta$ & 1.3 & \ref{employee_sales}(b) \\  \hline 
$<s|l,k>_{0.5} \propto l^{{\gamma_{s|l,k}}^{(l)}}$ & ${\gamma_{s|l,k}}^{(l)}$ & $\alpha $  & 0.9 & \ref{real_1d}(a) \\ \hline  
$<s|l,k>_{0.5} \propto k^{{\gamma_{s|l,k}}^{(k)}}$ & ${\gamma_{s|l,k}}^{(k)}$ & $\beta $ & 0.4 & \ref{real_1d}(b) \\ \hline
$<l|k,s>_{0.5} \propto k^{{\gamma_{l|k,s}}^{(k)}}$ & ${\gamma_{l|k,s}}^{(k)}$ & $ \frac{-(\alpha+\beta) \cdot \alpha}{\alpha^2+\sigma_A^2/\sigma_B^2}+1$ & 0.1 & \ref{real_1d}(c) \\ \hline
$<l|k,s>_{0.5} \propto k^{{\gamma_{l|k,s}}^{(s)}}$ & ${\gamma_{l|k,s}}^{(s)}$ & $\frac{\alpha}{\alpha^2+\sigma_A^2/\sigma_B^2}$ & 0.7 &  \ref{real_1d}(d) \\ \hline \hline
\end{tabular}
\caption{Summary of scaling exponents.}
\label{table1}
\end{table}

Next, we investigate how the scaling exponents depend on the magnitude fluctuations.
Here，the analytical calculation is done by approximating the distributions of $A$ and $B$ by log-normal distributions $\phi'(A;\mu_A$,$\sigma_A)$, $\phi'(B;\mu_B,\sigma_B)$ and $K$ by the Pareto distribution $q(K;\lambda,k_{m})$, where
\begin{equation}
\phi'(x;\mu,\sigma)=\frac{1}{\sqrt{2\pi}\sigma x}\exp \left(-\frac{{(\ln{(x)}-\mu)}^2}{2\sigma^2}\right) \quad (0 < x <\infty) \end{equation}
\begin{equation}
q(x;\lambda,x_{m})=\frac{\lambda x_m^{\lambda}}{x^{\lambda+1}} \quad (x_{m} \leq x <\infty). \label{K} 
\end{equation}
The real data is approximated at best with the set of parameters; $\mu_A=9.7$, $\sigma_A=0.88$, $\mu_B=0.72$ and $\sigma_B=1.2$, $\lambda=1.3$ and $k_m=3$. 
In this study, we refer to the set of these values as the best parameter set.
Here, $\mu_A$ and $\mu_B$ are estimated by the mean of the actual values $A$ and $B$, $\sigma_A$ and $\sigma_B$ are estimated by the standard deviation of the data. 
The quantities of $\lambda$ and $k_m$ are determined by the fitting of Eq. (\ref{K}) to the real data shown in Fig. \ref{sed}(a) by the green-dash-dotted lines. 
From Fig. \ref{theory}, we see that the central part of the 
actual distributions is reasonably approximated by these lognormal distributions for $A$ and $B$. 
However, significant disagreement occurs for the tail parts. 
In addition, for $K$, the tail part of the empirical distribution (i.e., above $k_m$) is well approximated by the above-mentioned Pareto distribution, which is shown by the green dash-dotted lines in Fig. \ref{sed}(a). 
\par 
\subsubsection{Correlations among three variables}
\label{sec:three}
Here, we discuss the relation between $<s|k,l>_{0.5}$, $<l|k,s>_{0.5}$ and $<k|l,s>_{0.5}$. If $A$ and $B$ follow a log-normal distributions and $K$ follows the Pareto distribution, we can calculate these values rigorously. The details of the derivation are given in Appendix \ref{sec:A2d}. 
In this section, we give only the results. \par
$<s|k,l>_{0.5}$ and $<l|k,s>_{0.5}$ can be written as
\begin{eqnarray}
&<&s|k,l>_{0.5} \propto  l^{\alpha} \cdot k^{\beta} \label{s_kl} \\
&<&l|k,s>_{0.5} \propto k^{-\nu \kappa_l+1}\cdot s^{\kappa_l},   \label{l_ks}
\end{eqnarray}
where $\nu=\alpha+\beta$ and $\kappa_l=\frac{\alpha}{\alpha^2+\sigma_A^2/\sigma_B^2}$. 
These equations agree well with the real data [see Figs. \ref{shuff_2d}(d) and (e)].
Here, the scaling indices for the conditional scaling relations are given by the model's parameters as $\gamma_{s|k,l}^{(k)}=\beta$, $\gamma_{s|k,l}^{(l)}=\alpha$, $\gamma_{l|k,s}^{(k)}=-\frac{(\alpha+\beta) \cdot \alpha}{\alpha^2+\sigma_A^2/\sigma_B^2}+1$ and $\gamma_{l|k,s}^{(s)}=\frac{\alpha}{\alpha^2+\sigma_A^2/\sigma_B^2}$. 
For the best parameter set, $\gamma_{s|k,l}^{(k)}=0.4$, $\gamma_{s|k,l}^{(l)}=0.9$, $\gamma_{l|k,s}^{(k)}=0.1$ and $\gamma_{l|k,s}^{(s)}=0.7$, which agrees with empirical scaling indices (see Table 1). 
Note that Eq. (\ref{l_ks}) implies that the scaling exponent of $<l|k,s>_{0.5}$  depends on the magnitude of the fluctuations of $A$ and $B$. 
For example, in the limit $\sigma_A^2 \to 0$, we have $<l|k,s>_{0.5} \propto k^{-\beta/\alpha}\cdot s^{1/\alpha}$, which  corresponds to the analytical solution of $s=A \cdot l^\alpha \cdot k^\beta$ given by Eq. (\ref{model3}) with respect to $l$ neglecting the fluctuation. 
Conversely, for the $\sigma_B^2 \to 0$, $<l|k,s>_{0.5} \propto k$, which corresponds to the solution of $l=B \cdot  k $ given by Eq. (\ref{model2}).  \par

From the rigorous formula of $<k|l,s>$ given by Eq. (\ref{kls_a}) and in the case of best parameter set，we can get the following the scaling law for $z \to -\infty$:
\begin{equation}
<k|l,s>_{0.5} \propto l^{-\nu\kappa_k+1}  \cdot s^{\kappa_k}, \label{k_ls_p}
\end{equation}
where $\kappa_k=\beta/(\beta^2+\sigma_A^2/\sigma_B^2)$.

In the central region, the theoretical curves roughly agree with actual curves; however, they disagree 
near the extremities. 
Comparing Fig. \ref{shuff_2d}(c) with Fig. \ref{shuff_2d}(f), we see that the cause for disagreements around the edges of the contour lines comes from the deviation in the tail portion of the distributions of $A$ and $B$ (shown in Fig. \ref{theory}). 
\subsubsection{Correlations between two variables}
\label{sec:two}
Here, we calculate the conditional distributions for two variables theoretically based on the log-normal model. The details of derivation are given in Appendix \ref{sec:A1d}. \par

From Eqs. (\ref{l_k_q_a}) and (\ref{s_k_q_a}), the conditional percentiles of $l$ given by $k$ 
and the conditional percentiles of $s$ given $k$ can be written as: 
\begin{equation}
<l|k>_{q} \propto k. \quad (0 \leq q \leq 1) \label{th_lk_q}
\end{equation}
\begin{equation}
<s|k>_{q} \propto k^{\nu} \quad (0 \leq q \leq 1), \label{th_sk_q}
\end{equation}
which corresponds to the empirical equations Eq. (\ref{l_k_kika}) and equation Eq. (\ref{s_k_kika}) respectively.
Thus, $\gamma_{l|k}=1$ and $\gamma_{s|k}=\nu=\alpha+\beta=1.3$ for the best parameter set.
Similarly, from Eq. (\ref{s_l_q_a}), we have the conditional percentiles of $s$ given $l$, 

\begin{eqnarray}
<s|l>_{q} \propto l^{\nu}
\quad (l \to \infty; \quad \sigma_A^2/\sigma_B^2 > \alpha \beta ), \label{th_sl_q}
\end{eqnarray}
where $0 \leq $q$ \leq 1$. This equation corresponds to the empirical equation Eq. (\ref{s_l_kika}), namely, $\gamma_{s|l}=\nu=\alpha+\beta=1.3$. 
\par 
We can also analytically calculate the conditional distributions, $P(l|k)$, $P(s|k)$ and $P(l|k)$. 
From Fig. \ref{syuuhen2}, we see that, except for the tail portions, the empirical curves plotted as black lines agree well with the green dash-dotted line, which ensures the validity of Eqs. (\ref{l_k_p_a}), (\ref{s_k_p_a}) and (\ref{s_l_p_a}). 
Note that the discrepancies are again because of the deviations in the tail portions of the distributions of coefficients for $A$ and $B$. \par
\subsubsection{Marginal distributions}
\label{sec:one}
Finally, we calculated the conditional distributions for the marginal distributions. The details of the derivation are given in Appendix \ref{sec:A0d}. \par
 The marginal distribution of $k$ is given by the distribution of $K$, $q(K;\lambda,k_{m})$.
 With $k$ is given by Eq. (\ref{model1}), we have the following power law distribution:
\begin{equation}
P(k) \propto k^{-\lambda-1} \quad (k_m \leq k < \infty) \label{pdf_k}.
\end{equation}
Thus, $\zeta_k=\lambda$, which takes $1.3$ for the best parameter set.
\par 
The asymptotic behavior of the marginal distributions of $l$ and $s$ are derived as:
\begin{equation}
P(l) \propto l^{-\lambda-1} \quad (l \to \infty). \label{pdf_l}
\end{equation}
\begin{equation}
P(s) \propto s^{-\frac{\lambda}{\alpha+\beta}-1}  \quad (s \to \infty). \label{pdf_s}
\end{equation} 
Thus, for the best parameter set, $\zeta_{l}=\lambda$, which takes $1.3$, and $\zeta_{s}=\lambda/(\alpha+\beta)$ which takes $1.0$. 
These equations correspond to the empirical PDF given by Eq. (\ref{power}).  \par
The green dash-dotted lines in Figs. \ref{sed} (a)-(c) are the theoretical curves given by Eqs. (\ref{pk_a}), (\ref{pl_a}) and (\ref{ps_a}).  
These figures show that the empirical distribution are closely fit by the theoretical curves. \par 
Table \ref{table1} summarizes the scaling exponents mentioned in Sec. \ref{sec:observation} derived from empirical observations and the corresponding theoretical exponents discussed in Sec. \ref{sec:theory} 
 
\section{Discussion and Conclusion}
In this study, we analyzed the scaling behavior of scale indicators for Japanese firms. In particular, we focused three basic scale indicators: sales (flow value), number of employees (stock value), and number of business partners (business relation). First, by analyzing the financial data of about 500,000 Japanese firms, we established the following relations:
\begin{itemize}
\item[(i)] The conditional percentiles scale with the exponent about 1.0 for number of employees based on degrees, with exponent about 1.3 for sales based on degrees, and with exponent about 1.3 for sales based on the number of employees; 
\item[(ii)] Corresponding conditional distribution functions converge into a unique scaling function, through the scaling relations of the conditional medians,  respectively; 
\item[(iii)]New scaling relations appear between three variables, such as the scalings of conditional median of sales based on the numbers of business partners and employees.
\end{itemize} \par
Second, we introduced simple stochastic models that reproduce all empirical scaling relations consistently, and we derived the nontrivial relation between scalings indices and fluctuations. To provide a consistent explanation of these three-body scaling relations, we show that it is necessary to consider the effects of fluctuations in coefficients. In other words, scaling indices depend on the magnitude of the fluctuations. It is interesting that for two-body relations, which have been well cultivated, the fluctuations do not modulate the exponents. 
To clarify such an effect on the allometric scaling relations, a more in-depth study is required into situations involving more than three variables. \par

Regarding the scaling of the metabolic rate of mammals, the geometric structure of a vessel network has been shown to explain the allometric properties. Similarly, we can pose a basic question; namely, can we explain our empirical scaling relations from the network structure of the interfirm trading relation? In our recent study, we showed that the scaling of sales based on degree and with exponent $1.3$ and the power law distribution of sales with the exponent 1 are explained by the transport of money through the interfirm trading network \cite{Watanabe2011}. 
Moreover, this transport model explains the scaling of sales based on employees with exponent $1.3$. 
However, in the present form, this transport model cannot reproduce all the scaling relations for the three variables. 
It is our task in the near future to pursue the network model, so that the key coefficients $A$ and $B$ in Eqs. (\ref{model1})-(\ref{model3}) can be estimated by the information of the network structure. 
We can also associate these properties with the interfirm trading network. 
A detail survey of along these lines will be reported in a future presentation.

\begin{acknowledgments}
We thank the Research Institute of Economy, Trade and industry 
(RIETI) for allowing us to use the TSR data. This work is partly supported by Grantin-
Aid for JSPS Fellows Grant No. 219685 (H.W.) and Grant-in-Aid for Scientiist
Research No. 22656025 (M.T.) from JSPS.
\end{acknowledgments}
\appendix

\section{Conditional medians, $<s|k,l>_{0.5}$, $<l|k,s>_{0.5}$, and $<k|l,s>_{0.5}$} 
\label{sec:A2d}
Here, we calculate the median of s for given $l$ and $k$, $<s|l,k>_{0.5}$.
Taking the logarithm, Eqs. (\ref{model1})-(\ref{model3}) can transform into 
\begin{equation}
k'= K' \label{lmodel1}
\end{equation}
\begin{equation}
l'=B'+k' \label{lmodel2}
\end{equation}
\begin{equation}
s'=A'+\alpha l'+\beta k' \label{lmodel3}
\end{equation}
where $s'=\log(s)$, $k'=\log(k)$, $l'=\log(l)$, 
$A'=\log(A)$, $B'=\log(B)$ and $K'=\log(K)$. 
Thus, the PDF of $A'$ is $\phi(A';\mu_A,\sigma_A)$, 
the PDF of $B'$ is $\phi(B';\mu_B,\sigma_B)$ and the PDF of $K'$ is $ \lambda \exp(\lambda \cdot (K'-\log(km)))$.
Here, $\phi(x, \mu,\sigma)$ is the PDF of the normal distribution whose mean is $\mu$ and standard deviation is $\sigma$.
Because $A'$ obeys the normal distribution, the conditional mean of logarithmic sales $<s'|k',l'>$ is  
\begin{equation}
<s'|k',l'>=<A'>+\alpha l'+\beta k'=\mu_A+\alpha l'+\beta k'.
\end{equation}
Thus, we get
\begin{eqnarray}
<s|k,l>_{0.5}=\exp{(<s'|k',l'>)}  
=\exp(\mu_A) l^{\alpha} k^{\beta}, \label{skh}
\end{eqnarray}
where we use a property of the lognormal distribution, namely, if $x$ has the lognormal PDF $\phi'(x,\mu,\sigma)$, then the median of $x$ is $\exp(\mu)$. This equation is Eq. (\ref{s_kl}) in Sec. \ref{sec:three}. \par

Similarly, we calculate the value of $<l|k,s>_{0.5}$．
From Eqs. (\ref{lmodel2}) and (\ref{lmodel3}), we have  
\begin{equation}
s'=A'+\beta k'+ \alpha (B'+k')=\nu k'+c,  \label{l_c}
\end{equation}
where $c=A'+\alpha B'$ and $\nu=\alpha+\beta$. \\
Here, $c$ is fixed for a given $k$ and $s$ is determined. 
Therefore, the condition by $k$ and $s$ is equivalent to the condition by 
$c=s'-\nu k'$:
\begin{equation}
<B'|k',s'>=<B'|c>. \label{l_c_2}
\end{equation}
From Bayes' theorem，the conditional probability of $B'$ is estimated using 
\begin{eqnarray}
P(B'|c) &\propto& P(c|B') P(B') \nonumber \\
&\propto& \phi(c;\mu_A+\alpha B',\sigma_A) \cdot \phi(B';\mu_B,\sigma_B) \nonumber \\
& \propto & \phi(B';\mu_{B'|c},\sigma_{B'|c}),
\end{eqnarray}
where $\mu_{B'|c}=\kappa_lc+\tau_l$, $\sigma_{B'|c}=\sigma_A \sqrt{\kappa_l/\alpha}$
,$\kappa_l=\alpha/(\alpha^2+\sigma_A^2/\sigma_B^2)$ and $\tau_l= \kappa_l/\alpha \cdot (- \alpha \mu_A+\sigma_A^2/\sigma_B^2 \cdot  \mu_B)$.
Thus, we have $<B'|c>=\kappa_lc+\tau_l$.
We take the conditional mean of Eq. (\ref{lmodel2}) and substitute it into Eq. (\ref{l_c_2}), which gives 
\begin{eqnarray}
<l'|k',s'>=k+<B'|k',s'>  
=(-\nu \kappa_l+1)k'+\kappa_ls'+\tau_l. \nonumber \\ 
\end{eqnarray}
Therefore，
\begin{eqnarray}
<l|k,s>_{0.5}=\exp(<l'|k',s'>_{0.5}) \propto k^{-\nu \kappa_l+1} s^{\kappa_l}. 
\end{eqnarray}
This equation gives Eq. (\ref{l_ks}) in Sec. (\ref{sec:three}).
Here, we use a property of the lognormal distribution; that is, if $x$ has the lognormal PDF $\phi'(x,\mu,\sigma)$, then the median of $x$ is given by $\exp(\mu)$. \par

We also calcualte $<k|l,s>_{0.5}$, Because A' and B' obey the normal distributions [from Eqs. (\ref{lmodel2}) and (\ref{lmodel3})], 
the conditional probability of $l'$ and $s'$ for given $k$' is written as: 
\begin{equation}
P(l',s'|k')=f(l',s';\mu_{s'|k'},\mu_{l'|k'},\sigma_{s'|k'},\sigma_{l'|k'},\rho)
\end{equation}
where 
\begin{eqnarray}
\mu_{s'|k'}&=&\nu k'+\mu_A+\alpha \mu_B, \nonumber \\
\mu_{l'|k'}&=&k'+\mu_B, \nonumber \\
\sigma_{s'|k'}^2&=&\sigma_A^2+\alpha^2\sigma_B^2, \nonumber \\
\sigma_{l'|k'}^2&=&\sigma_B^2, \nonumber \\
\rho&=& \alpha \frac{\sigma_{l'|k'}}{\sigma_{s'|k'}}, \nonumber 
\end{eqnarray}
and $f$ is the PDF of the multivariate normal distribution:
\begin{eqnarray}
&&f(x,y;\mu_x,\mu_y,\sigma_x,\sigma_y,\rho) \nonumber \\  
&=&\frac{1}{2 \pi \sigma_x \sigma_y \sqrt{1-\rho^2}}\exp(-\frac{1}{2(1-\rho^2)}\cdot ( \frac{(x-\mu_x)^2}{\sigma_x^2} \nonumber \\ 
&+&\frac{(y-\mu_y)^2}{\sigma_y^2}-\frac{2\rho \cdot (x-\mu_x)(y-\mu_y)}{\sigma_x\sigma_y})). \nonumber
\end{eqnarray}
Applying the Bayes' theorem, we have the following relation: 
\begin{eqnarray}
P(k'|l',s') &\propto& P(l,s|k')P(k') \nonumber \\
&\propto& \phi(k';\mu_{k'|l',s'},\sigma_{k'|l',s'}), 
\end{eqnarray}
where 
\begin{eqnarray}
\mu_{k'|l',s'}&=&(-\nu\kappa_k+1)l'+\kappa_k s'+\tau_k, \nonumber \\
{\sigma_{k'|l',s'}} &=& \sqrt{(1-\rho^2)\cdot \kappa_k/\kappa_l \cdot \alpha/\beta} \cdot \sigma_B,  \nonumber \\
\kappa_k&=&\beta/(\beta^2+\sigma_A^2/\sigma_B^2), \nonumber \\ 
\tau_k&=&-(\beta \mu_A+\mu_B \sigma_A^2/\sigma_B^2)\cdot \kappa_k/\beta - \lambda \sigma_{k'|l',s'}^2 , \nonumber 
\end{eqnarray}
and the support of $P(k'|l',s')$ is $\log(k_m) \leq k' < \infty$. 
In other words, $k'|l',s'$ obeys the truncated normal distribution with the following parameters; the mean is $\mu_{k'|l',s'}$, 
the standard deviation is $\sigma_{k'|l',s'}$, the minimum value is $\log(k_{m})$ and the maximum value is $\infty$.
Applying a general formula of the median of a truncated normal distribution, we get 
\begin{eqnarray}
&<&k'|l',s'>_{0.5} \nonumber \\
&=& \sigma_{k'|l',s'} \cdot \Phi_0^{-1}\left(\Phi_0(z)+\frac{1}{2}(1-\Phi_0(z))\right)+\mu_{k'|l',s'}. \nonumber \\
\label{kls_a}
\end{eqnarray}
where  
\begin{equation}
z=\frac{\log(k_m)-\mu_{k'|l',s'}}{\sigma_{k'|l',s'}} \nonumber 
\end{equation}
and, 
\begin{equation}
\Phi_0(x)=\frac{1}{\sqrt{2\pi}}\int^{x}_{-\infty}\exp(-\frac{t^2}{2})dt. \nonumber 
\end{equation}
Therefore, we have the following scaling relation:
\begin{equation}
<k|l,s>_{0.5} \propto l^{-\nu\kappa_k+1}  \cdot s^{\kappa_k} \quad (z \to -\infty). \label{k_ls_pp}
\end{equation}
This equation is Eq. (\ref{k_ls_p}) in Sec. \ref{sec:three}. 
 
\section{Conditional distributions of $l|k$, $s|k$ and $s|l$}
\label{sec:A1d}
We now calculate the conditional distribution for two variables. 
Let us consider the conditional random variable $l$ for given $k$, $l|k$. 
From Eq. (\ref{lmodel2}), we get the conditional distribution of $l'$ given by $k'$ as 
\begin{equation}
P(l'|k')=\phi(l';\mu_B+k',\sigma_B). \label{v2_l_k}
\end{equation}
Thus, the distribution of $l|k$ is given by 
\begin{equation}
P(l|k)=\phi'(l;\mu_B+\log(k),\sigma_B). \label{l_k_p_a}
\end{equation}
Therefore，the conditional percentile $100q$ of $l|k$ is written as
\begin{equation}
<l|k>_{q}=\exp(<l'|k'>_q) \propto \exp(\log(k)) \propto k. \label{l_k_q_a}
\end{equation} 
Eq. (\ref{l_k_q_a}) corresponds to Eq. (\ref{th_lk_q}) in Sec. \ref{sec:two}. \par

Next, we discuss $s|k$.
From Eqs. (\ref{lmodel1}) and (\ref{lmodel2}), 
 we obtain 
\begin{equation}
P(s'|k')=\phi(s';\mu_A+\beta \mu_B+(\alpha+\beta)k',\sigma_{s'|k'}), 
\end{equation}
where $\sigma_{s'|k'}=\sqrt{\sigma_A^2+\alpha^2\sigma_B^2}$. \\
Accordingly，the distribution of $s|k$ is 
\begin{equation}
P(s|k)=\phi'(s';\mu_A+\beta \mu_B+\nu \log(k),\sigma_{s'|k'}) \label{s_k_p_a}
\end{equation}
Thus, the conditional percentile $100q$ of $s|k$ is written as:
\begin{equation}
<s|k>_{q}=\exp(<s'|k'>_q) \propto \exp(\nu \log(k)) \propto k^{\nu}. \label{s_k_q_a}
\end{equation}
This equation 
is Eq. (\ref{th_sk_q}) in Sec. \ref{sec:two}. \par
Similarly, we consider $s|l$．
From Eq. (\ref{v2_l_k}) and Bayes' theorem: 
\begin{equation}
P(k'|l')\propto P(l'|k')P(k') \propto \phi(k';l'-\mu_B-\lambda\sigma_B^2,\sigma_B^2), 
\end{equation}
where the support of this distribution is $\log(k_m) \leq k' < \infty$.
Here, from Eq. (\ref{lmodel3}), we have the following relation: 
\begin{equation}
s'|l'=A'+\alpha l'+\beta k'|l'.
\end{equation}
Note that $A$ and $k'|l'$ are independent of each other, so by 
taking a convolution of the probability distribution function of $A$ and $k'|l'$, we arrive at the following relation:  
\begin{eqnarray}
P(s'|l') &\propto&\int^{\infty}_{\beta \log(k_m)}\phi(s'-x;\mu_A+\alpha l',\sigma_A) \nonumber \\ 
&\cdot&\phi(x;\beta(l'-\mu_B-\lambda\sigma_B^2),\beta \sigma_B) dx \nonumber \\
&\propto&\phi(s';M_1'(l'),S_1) \nonumber \\
&\cdot& \left\{(1-\Phi(\beta \log(k_m);M_2'(s',l'),S_2) \right\}, 
\end{eqnarray}
where
\begin{equation}
M_1'(l')=\nu l'+\mu_A+\beta(-\mu_B-\lambda\sigma_B^2), \nonumber
\end{equation}
\begin{equation}
S_1=\sqrt{\sigma_A^2+\beta^2\sigma_B^2}, \nonumber 
\end{equation}
\begin{eqnarray}
M_2'(s',l')&=& \kappa_k \left(-\alpha \beta+\frac{\sigma_A^2}{\sigma_B^2} \right)l'+\beta \kappa_k s' \nonumber \\
&-&\kappa_k\left( \beta \mu_A+\frac{\sigma_A^2}{\sigma_B^2} (\mu_B+ \lambda \sigma_B^2) \right), \nonumber 
\end{eqnarray}
\begin{equation}
S_2={\left(\frac{1}{\sigma_A^2}+\frac{1}{\beta^2\sigma_B^2}\right)}^{-\frac{1}{2}} \nonumber,
\end{equation}
and 
\begin{equation}
\Phi(x;\mu,\sigma)=\int^{x}_{-\infty}\phi(t;\mu,\sigma)dt. \nonumber
\end{equation}
Consequently，the distribution of $s|l$ is estimated by using  
\begin{eqnarray}
P(s|l)&\propto&\frac{1}{s}\phi(\log(s);M_1(l),S_1) \nonumber  \\ 
&\cdot& \left\{ 1-\Phi(\beta \log(k_m);M_2(s,l),S_2) \right\},  \nonumber \\ \label{s_l_p_a}
\end{eqnarray}
where $M_1(l)=M_1'(\log(l))$ and $M_2(s)=M_2'(\log(s))$.
 \par
If $\sigma_A^2/\sigma_B^2 > \alpha \beta$, 
we can get the following asymptotic behavior for $l \to \infty$:   
\begin{equation}
P(s|l) \propto \frac{1}{s}\phi(\log(s);M_1(l),S_1) \propto \phi'(s; M_1(l),S_1).
\end{equation}
Therefore, we have the following scaling relation: 
\begin{eqnarray}
<s|l>_{q} \propto \exp{(\nu\log(l))} &\propto& l^{\nu} \label{s_l_q_a}.
\quad (l \to \infty).
\end{eqnarray}
This equation is Eq. (\ref{th_sl_q}) in Sec. \ref{sec:two}.

\section{Marginal distributions of $k$, $l$ and $s$}
\label{sec:A0d}
Finally, we calculate the marginal distributions of $k$, $l$, and $s$. 
Because of the definition of $k$, the marginal distribution of $k$ is the same as the distribution of $K$.
Therefore,   
\begin{equation}
P(k)=\frac{\lambda k_m^\lambda}{k_m^{\lambda+1}}\propto k^{-\lambda-1} \quad (k_m \leq k < \infty). \label{pk_a}
\end{equation} 
This equation corresponds to Eq. (\ref{pdf_k}) in Sec. \ref{sec:one}.
\par
Next, we consider the marginal distribution of $l$. 
Eqs. (\ref{lmodel1}) and (\ref{lmodel2}) mean that $l$ is the sum of two independent random variables: $K'$ and $B'$. 
Therefore, taking the convolution of the PDF of $B'$ and $K'$, we have 
\begin{eqnarray}
P(l') &=&\int^{\infty}_{\log(k_m)} \phi(l'-x;\mu_B,\sigma_B) \lambda \exp(-\lambda(x-x_m))dx  \nonumber \\
&\propto& \exp (-l' \lambda) \left\{1-\Phi(\log(k_m);l' -\mu_B-\sigma_B^2\lambda,\sigma_B) \right \} \nonumber. \\ 
\end{eqnarray}
Thus, the marginal distribution of $l$ is  
\begin{eqnarray}
P(l)&\propto& l^{-\lambda-1}  \left\{1-\Phi(\log(k_m);\log(l)-\mu_B-\sigma_B^2\lambda,\sigma_B) \right \}  \nonumber. \\ \label{pl_a} \end{eqnarray}
Because $\Phi(\log(k_m);\log(l)-\mu_A-\sigma_B^2\lambda,\sigma_B) \to 0$
for $l \to \infty$, we have following asymptotic behavior:
\begin{equation} 
P(l) \propto l^{-\lambda-1}.
\end{equation}
This is Eq. (\ref{pdf_l}) in Sec. \ref{sec:one}.
\par
Similarly, we calculate the marginal distribution of $s$． 
From Eqs. (\ref{lmodel1}), (\ref{lmodel2}) and (\ref{lmodel3}), 
we get 
\begin{equation}
s'=\nu K'+A'+\alpha B'.
\end{equation}
Then, its PDF is obtained by taking the convolution $A+\alpha B$ with the PDF $\phi(x,\mu_A+\alpha\mu_B,\sigma_{s'|k'})$ and $\nu K$ with the PDF $ \lambda/\nu \exp(\lambda/\nu (x-x_m))$:
\begin{eqnarray}
P(s') &=&\int^{\infty}_{\log(k_m)} \phi(s'-x;\mu_A+ \alpha \mu_B,\sigma_{s'|k'})
 \nonumber \\
&\cdot& \frac{\lambda}{\nu} \exp(-\frac{\lambda}{\nu}(x-x_m))dx \nonumber \\
&\propto& \exp (-\frac{s'\lambda}{\nu}) \nonumber \\ 
&\cdot & \left\{1-\Phi(\log(k_m);s'-\mu_A- \alpha \mu_B-\frac{\lambda \sigma_{s'|k'}^2}{\nu},\sigma_{s'|k'}) \right\}. \nonumber \\
\end{eqnarray}
Thus, the marginal distribution of $s$ is 
\begin{eqnarray}
P(s)&\propto& s^{-\frac{\lambda}{\alpha+\beta}-1} \nonumber \\ 
&\cdot & \left\{1-\Phi(k_m;\log(s)-\mu_A- \alpha \mu_B-\frac{\lambda \sigma_{s'|k'}^2}{\nu} , \sigma_{s'|k'}  \right\}.
\nonumber \\ \label{ps_a}
\end{eqnarray}
For $s \to \infty$, 
we get the following asymptotic behavior: 
\begin{equation}
P(s) \propto s^{-\frac{\lambda}{\alpha+\beta}-1} \quad (s \to \infty).
\end{equation}
This is Eq. (\ref{pdf_s}) in Sec. \ref{sec:one}.
\par

\bibliography{scaling2.bib}

\begin{thebibliography}{10}

\bibitem{stahl1965organ}
W.~Stahl,
\newblock Science {\bf 150}, 1039 (1965).

\bibitem{Kleiber1947}
M.~Kleiber,
\newblock Physiol. Rev. {\bf 27}, 511 (1947).

\bibitem{West1997}
G.~B. West, J.~H. Brown, and B.~J. Enquist,
\newblock Science {\bf 276}, 122 (1997).

\bibitem{Bettencourt2007}
L.~Bettencourt, J.~Lobo, D.~Helbing, C.~K{\"u}hnert, and G.~B. West,
\newblock Proc. Natl. Acad. Sci. USA {\bf 104}, 7301 (2007).

\bibitem{Bettencourt2010}
L.~M.~A. Bettencourt, J.~Lobo, D.~Strumsky, and G.~B. West,
\newblock PLOS ONE {\bf 5}, e13541 (2010).

\bibitem{horta2010}
R.~Horta-Bern{\'u}s, M.~Rosas-Casals, and S.~Valverde,
\newblock in {\em COMPENG'10.}, pp. 49--51, Rome, Italy, 2010.

\bibitem{Samaniego2008}
H.~Samaniego and M.~E. Moses,
\newblock JTLU {\bf 1} (2008).

\bibitem{Zhang2010}
J.~Zhang and T.~Yu,
\newblock Physica A {\bf 389}, 4887 (2010).

\bibitem{Stanley1996}
M.~H.~R. Stanley {\em et~al.},
\newblock Nature (London) {\bf 379}, 804 (1996).

\bibitem{Labra2007}
F.~A. Labra, P.~A. Marquet, and F.~Bozinovic,
\newblock Proc. Natl. Acad. Sci. USA {\bf 104}, 10900 (2007).

\bibitem{fujiwara2004pareto}
Y.~Fujiwara, C.~Di~Guilmi, H.~Aoyama, M.~Gallegati, and W.~Souma,
\newblock Physica A {\bf 335}, 197 (2004).

\bibitem{fu2005growth}
D.~Fu {\em et~al.},
\newblock Proc. Natl. Acad. Sci. USA {\bf 102}, 18801 (2005).

\bibitem{plerou1999similarities}
V.~Plerou, L.~A.~N. Amaral, P.~Gopikrishnan, M.~Meyer, and H.~E. Stanley,
\newblock Nature (London) {\bf 400}, 433 (1999).

\bibitem{mendes2006scaling}
S.~Picoli~Jr, R.~S. Mendes, L.~C. Malacarne, and E.~K. Lenzi,
\newblock Europhys. Lett. {\bf 75}, 673 (2006).

\bibitem{mendes2005statistical}
S.~Picoli~jr, R.~S. Mendes, and L.~C. Malacarne,
\newblock Europhys. Lett. {\bf 72}, 865 (2005).

\bibitem{picoli2008universal}
S.~Picoli~Jr and R.~S. Mendes,
\newblock Phys. Rev. E {\bf 77}, 036105 (2008).

\bibitem{keitt1998dynamics}
T.~H. Keitt and H.~E. Stanley,
\newblock Nature (London) {\bf 393}, 257 (1998).

\bibitem{axtell2001zipf}
R.~L. Axtell,
\newblock Science {\bf 293}, 1818 (2001).

\bibitem{okuyama1999zipf}
K.~Okuyama, M.~Takayasu, and H.~Takayasu,
\newblock Physica A {\bf 269}, 125 (1999).

\bibitem{aoyama2000pareto}
H.~Aoyama {\em et~al.},
\newblock Fractals {\bf 8}, 293 (2000).

\bibitem{Aoyama2010}
H.~Aoyama, Y.~Fujiwara, and M.~Gallegati,
\newblock Micro-Macro Relation of Production-The Double Scaling Law for
  Statistical Physics of Economy, arXiv:1003.2321 .

\bibitem{Watanabe2011mega}
T.~Watanabe, T.~Mizuno, A.~Ishikawa, and H.~Fujimoto,
\newblock The Economic Review (Keizai Kenkyuu) {\bf 62}, 193 (2011).

\bibitem{Saito2007}
Y.~U. Saito, T.~Watanabe, and M.~Iwamura,
\newblock Physica A {\bf 383}, 158 (2007).

\bibitem{Ohnishi2009}
T.~Ohnishi, H.~Takayasu, and M.~Takayasu,
\newblock Prog. Theor. Phys. Suppl. {\bf 179}, 157 (2009).

\bibitem{Watanabe2011}
H.~Watanabe, H.~Takayasu, and M.~Takayasu,
\newblock New J. Phys. {\bf 14}, 043034 (2012).

\end{thebibliography}
\end{document}